\documentclass{aastex63}

\usepackage{graphicx}
\usepackage{wrapfig}
\usepackage{natbib}

\newcommand{\be}{\begin{equation}}
\newcommand{\ee}{\end{equation}}
\newcommand{\nn}{\mbox{} \nonumber \\ \mbox{} }
\newcommand{\ba}{\begin{eqnarray}}
\newcommand{\ea}{\end{eqnarray}}
\newcommand{\om}{\omega}
\newcommand{\Alfven}{Alfv\'{e}n }

\newcommand{\curl}{{\rm curl\, }}
\newcommand{\A}{{\bf A}}
\newcommand{\E}{{\bf E}}
\newcommand{\B}{{\bf B}}
\newcommand{\J}{{\bf J}}
\renewcommand{\v}{{\bf v}}
\renewcommand{\k}{{\bf k}}

\newcommand{\sech}{{\rm \,sech\,}}

\newcommand\eg{{\it{e.g.\ }}}
\newcommand\cf{{\it{cf.\ }}}

\newcommand{\Bf}{{magnetic field}}
\newcommand{\Bfs}{{magnetic fields}}
\newcommand{\Ef}{{electric  field}}
\newcommand{\Efs}{{electric fields}}

\newcommand{\NS}{neutron star}
\newcommand{\NSs}{{neutron stars}}
\newcommand{\EM}{electromagnetic}

\newcommand{\ms}{magnetosphere}
\newcommand{\mss}{magnetospheres}

\newcommand{\LC}{light cylinder}
\newcommand{\Lf}{Lorentz factor}

\begin{document}

\title{Coherent  emission in pulsars, magnetars  and Fast Radio Bursts:  reconnection-driven free electron laser}
\author{Maxim Lyutikov}
\affil{Department of Physics  and Astronomy, Purdue University,   525 Northwestern Avenue, West Lafayette, IN47907-2036, USA; lyutikov@purdue.edu}


   \begin{abstract} 
We develop a model of the generation of coherent radio emission in the Crab pulsar, magnetars and Fast Radio Bursts (FRBs). Emission is produced by a reconnection-generated beam of particles via a variant of Free Electron Laser (FEL) mechanism, operating in a weakly-turbulent, guide-field dominated plasma. We first consider nonlinear  Thomson scattering in a guide-field dominated regime, and apply to model to explain emission bands observed in Crab pulsar and in Fast Radio Bursts. We consider particle motion in a combined fields of the  \EM\
 wave and the \EM\  (Alfvenic) wiggler. Charge bunches, created  via a ponderomotive force,  Compton/Raman  scatter the wiggler field coherently. 
The model is both robust to the underlying plasma parameters and succeeds in reproducing a number of subtle observed features: (i) emission frequencies depend mostly on the length $\lambda_t$  of turbulence and the Lorentz factor of the reconnection generated beam, $\omega \sim \gamma_b^2 ( c/\lambda_t) $  - it is independent of the absolute value of the underlying magnetic field. (ii) The model explains both broadband emission and the presence of emission stripes, including multiple stripes observed in the High Frequency Interpulse of the Crab pulsar. (iii) The model reproduces correlated polarization properties: presence of narrow emission bands in the spectrum favors linear polarization, while broadband emission can have arbitrary polarization. (iv) The mechanism is robust to the momentum spread of the particle  in the beam.  We also discuss a model of wigglers as non-linear force-free Alfven solitons (light darts).
   \end{abstract}

\section{Introduction:  challenges of coherent emission in Crab, magnetars and Fast Radio Bursts}

Pulsar emission mechanisms remain illusive for more than half a century \citep{1971ApJ...170..463G,1975ApJ...196...51R,cr77,1988Ap&SS.146..205B,1992RSPTA.341..105M,1999ApJ...521..351M,Melrose00Review,1999MNRAS.305..338L,2017RvMPP...1....5M,2018PhyU...61..353B}. Until recently, the rotationally-driven paradigm was prevailing: rotating \Bfs\ generate parallel \Efs, that accelerate particles with  unstable distribution function; this eventually lead to production of coherent emission \citep{1975ApJ...196...51R,1977ApJ...217..227F,AronsScharlemann,Hibschman}

During the last years a new consensus about generation of pulsar radio emission emerged, especially in application to  Crab Main Pulse  and Interpulse (MP and IP): radio emission is reconnection-driven, not rotationally-driven (there are likely several types of emission mechanism operating in pulsars). 
Reconnection has been suggested as a source of coherent emission in radio pulsars \citep{Istomin},  magnetar \citep{2002ApJ...580L..65L,2006MNRAS.367.1594L}.
and FRBs \citep{2013arXiv1307.4924P,2020ApJ...897....1L,2017ApJ...838L..13L,2020arXiv200505093L}. The reconnection-driven generation of radio pulses  is becoming a dominant theoretical concept  for generation of Crab MP and IP \citep{2010ApJ...715.1282B,2012SSRv..173..341A,2014ApJ...795L..22C,2015MNRAS.448..606C,2016MNRAS.457.2401C,2017SSRv..207..111C,2019SCPMA..6279511W,2019MNRAS.483.1731L,2019MNRAS.487..952C,2019ApJ...876L...6P}.    It is not clear why/how Crab is different from most other pulsars, in that its radio emission is dominated by  the reconnection-driven, not rotationally-driven processes. 

One of the most interesting  property of Crab  radio emission is observations of numerous emission stripes in Crab High Frequency interpulse 
\citep{1999ApJ...522.1046M,eilek07,2016ApJ...833...47H,2016JPlPh..82c6302E}.
 \begin{figure}[h!]
\centering
\includegraphics[width=.35\textwidth]{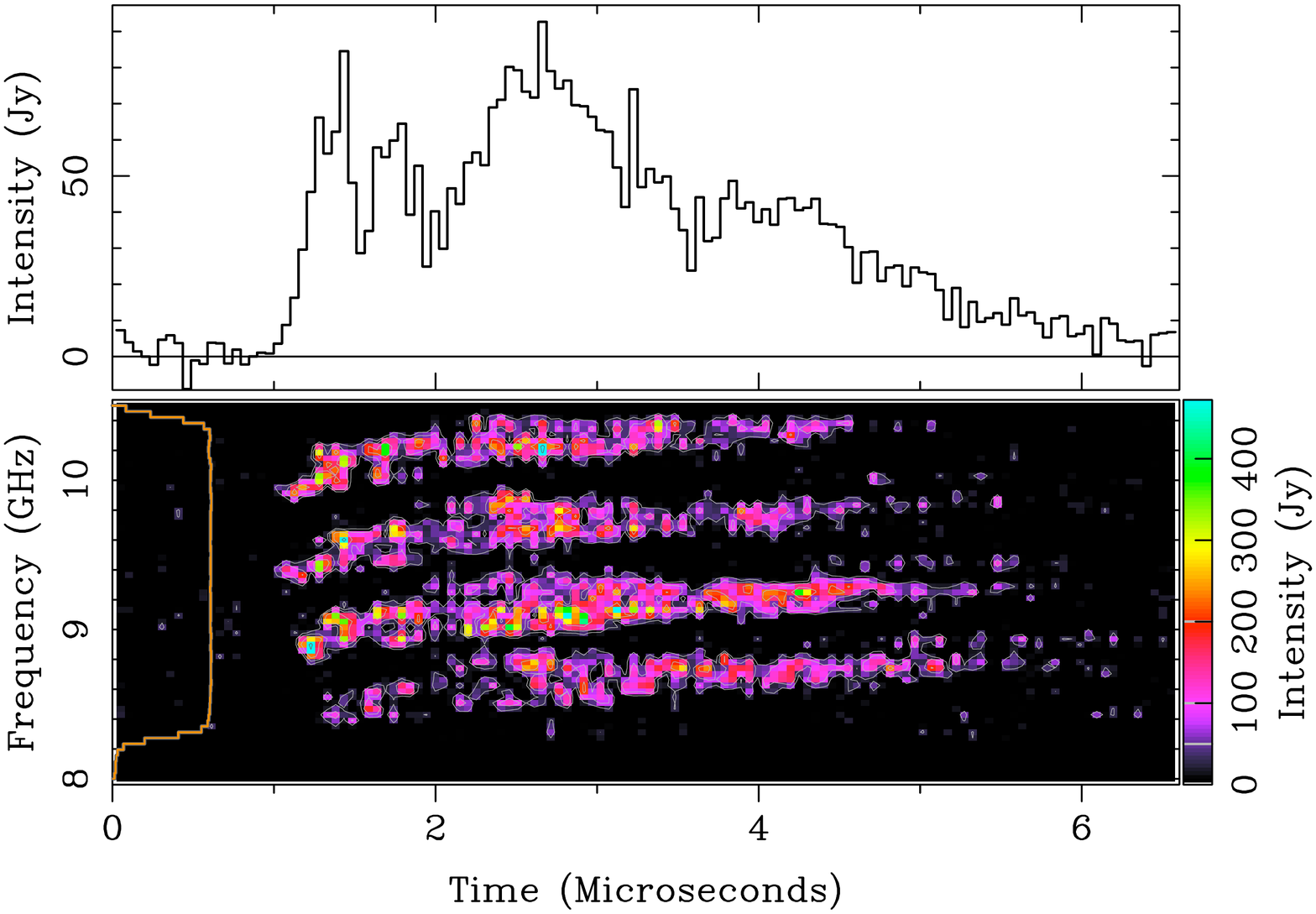} s
\includegraphics[width=.25\textwidth]{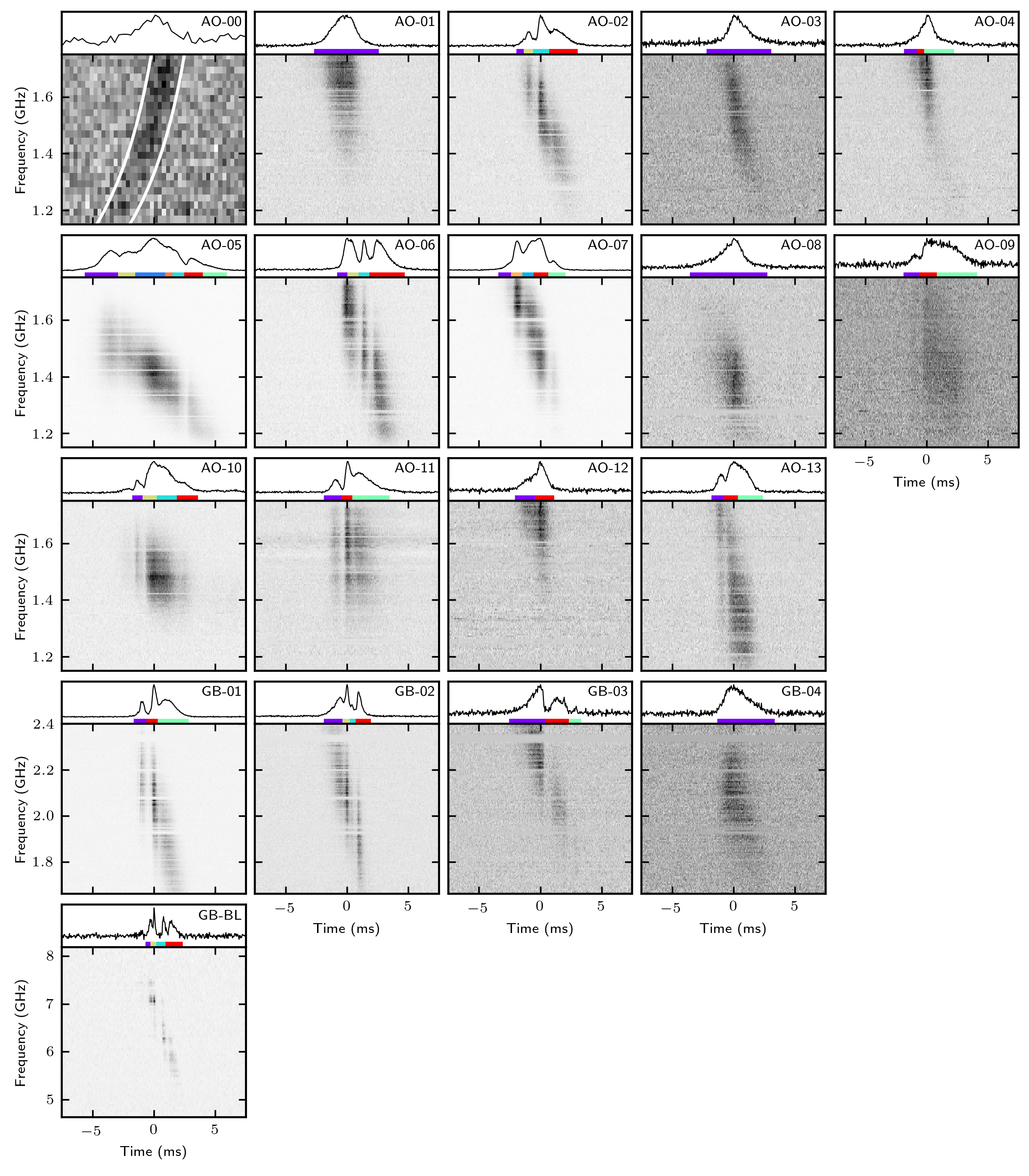}
\includegraphics[width=.35\textwidth]{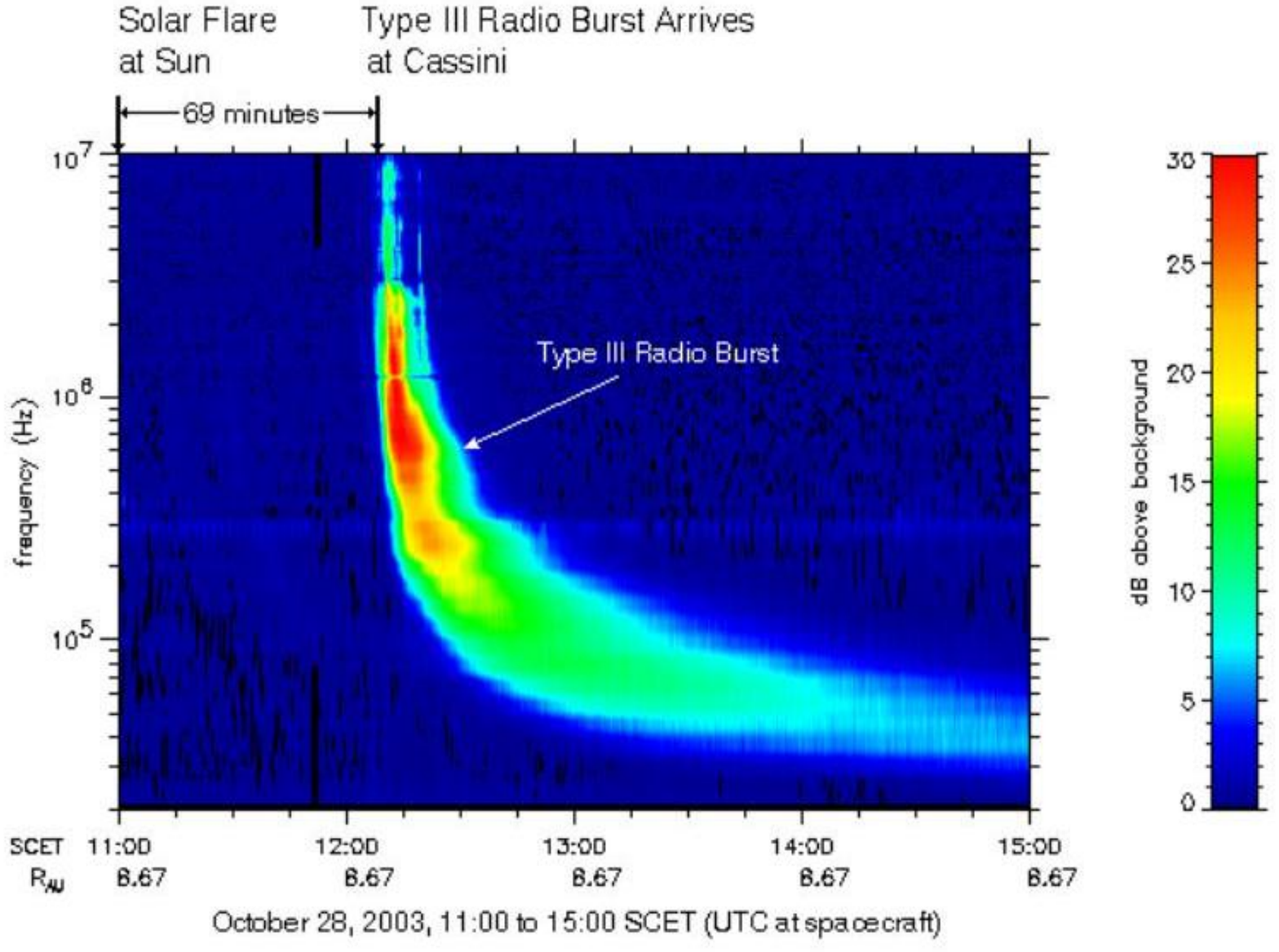}
\caption{Left panel: Example of narrow-band emission stripes in Crab High Frequency Interpulse \protect\cite{2016JPlPh..82c6302E}.  There is at least  more than a dozen bands extending from few to tens of GHz. Middle panel: narrow  drifting emission bands seen  in  FRB 121102  \protect\cite{2019ApJ...876L..23H}.
Right panel: Solar type-III radio burst showing similar downward frequency drifts to FRBs. 
}
\label{Fig4Right} 
\end{figure}
Emission is composed of at least a dozen relatively  narrow emission bands, almost equally spaced in frequency
(spacing increases with frequency at a rate of $6\%$).
 We consider  these stripes being the properties of the emission mechanism, not propagation \citep{2007MNRAS.381.1190L}. It is  one of the most intricate properties of pulsar radio emission - explaining it will be a strong argument in favor of any model of pulsar radio emission. 

Discoveries related to Fast Radio Bursts  \citep{2019A&ARv..27....4P,2019ARA&A..57..417C} renewed interest in production of coherent emission in compact objects. Simultaneous observations of radio and X-ray bursts \citep{2020Natur.587...54C,2020arXiv200511178R,2020Natur.587...59B,2020ApJ...898L..29M} establishes  magnetospheric origin of FRBs {\it loci} \citep{2020arXiv200505093L}. Observations of downward frequency drifts  \citep{2019ApJ...876L..23H,2019Natur.566..235C,2019arXiv190803507T,2019arXiv190611305J}
 are consistent with  the original prediction for magnetospheric origin of magnetar radio flares
\citep{2002ApJ...580L..65L}. 
Recent observations of PA swings \citep{2020arXiv201005800N}, originally predicted by \cite{2020ApJ...889..135L}, further confirm magnetospheric origin.   Also, the  long-term periodicity/binarity  \citep{2020Natur.582..351C,2020arXiv201208372P} is not related to the FRB mechanism, but is a propagation effect \citep{2020ApJ...893L..39L}. 

The observations of FRBs and magnetars  are consistent with the concept that radio and X-ray bursts are generated during reconnection events within  the \mss\ of  magnetars,  as suggested by  \cite{2002ApJ...580L..65L} \citep[see also][]{2020arXiv200505093L}.
 Conceptually,  magnetar  radio/X-ray flares  are  similar to Solar flares, initiated by the \Bf\ instabilities in the magnetars' \mss\ \citep[{\it not} the curst,][]{2012MNRAS.427.1574L,2015MNRAS.447.1407L}.
 By now, it seems, there is no escape in accepting the  magnetospheric origin of FRBs (X-ray bursts are magnetospheric events as demonstrated by the spin periodicity, radio leads X-rays, \cite{2020ApJ...898L..29M}) 

Previously a number of authors discussed FRBs as analogues of Crab Giant Pulses (GPs) \citep{2016MNRAS.462..941L,2016MNRAS.458L..19C,2016MNRAS.457..232C}.  The underlying assumption in those works was that Crab emission is rotationally powered. 
At that time  {\it three} types/mechanisms of radio emission thought to be operating in \NSs:  type-i rotationally-driven normal
pulses, exemplified by Crab precursor  (coming from opened field
lines, probably near the polar cap, having log-normal distribution in
fluxes), see \cite{1996ApJ...468..779M};  type-ii GPs, exemplified by Crab Main
Pulses and Interpulses (coming from outer \ms, near the last closed field
lines; having power-law distribution in fluxes \citep{1995ApJ...453..433L};
possibly with a special subset of supergiant pulses,
see \cite{2012ApJ...760...64M}; sometimes GPs show narrow spectral structure,
see \cite{2007ApJ...670..693H,2007MNRAS.381.1190L}. There can be several GP emission mechanisms, as the spectral properties of High Frequency IPs, and their rotational phase, differ from the MP and low frequency IP  \citep{1996ApJ...468..779M,2007ApJ...670..693H};  type-iii radio emission from magnetars
(coming from the region of close field lines, variable on secular times
scales and having very flat spectra) \cite{2006Natur.442..892C}.

\cite{2017ApJ...838L..13L} argued that the properties of the Repeater exclude rotationally-powered FRB emission.  The conclusion that FRBs cannot be rotationally-driven still stands. But  realization that Crab's GPs are not rotationally, but reconnection-powered is consistent with FRBs being analogues of {\it reconnection-powered} Crab GPs.

There is thus a clear uniting feature between Crabs' radio emission,  FRBs, and  Solar   type-III radio bursts  \citep{1963ARA&A...1..291W,2002ApJ...580L..65L,2020ApJ...889..135L,1985srph.book..333N,2020ApJ...889..135L}:
presence of narrow emission bands.
We know that Solar   radio bursts are reconnection-driven. {\it In the  present paper we develop a reconnection-driven model of pulsar's (Crab's in particular), magnetars'  and FRBs' radio emission.}

\section{Model in a nutshell }

We start with an assumption that \Bf\ lines are perturbed, \eg\ by a packet of \Alfven waves.
 \Alfven waves  are of low frequency,  much smaller than the beam plasma frequency $\om_{p,b}$ and with wavelength (somewhat) smaller than the local distance to center of the star.

In highly magnetized  force-free plasma \Alfven waves propagating along the \Bf\ are nearly luminal, $v_A \approx c$. 
In the setting of \NS\ \mss\ the difference between $v_A$ and the speed of light is much smaller than $1/\gamma^2$. 
Interaction of the beam with such waves will then resemble EM wiggler field.

 Let their wave number be $k_{w,b} $ (superscript ${(l)}$ implies that the quantity is measured in the lab frame.)
The wave  length $\lambda   = 2\pi/k_{w,b} $ is of the order of the local distance to the star,  $\lambda   \sim r$.
 
 A (reconnection-generated) beam of particles with \Lf\ $\gamma_b$ propagates along the wiggled \Bf\ in a direction opposite to the direction of \Alfven  waves. In the frame of the beam the waves are seen with $k_{w,b}= 2 \gamma_b k_{w} $. 
 
 In addition there is an \EM\ of frequency $\om$ wave propagating {\it along} with the beam. In the beam frame its wave number is reduced by $2 \gamma_b$. {\it In the beam frame the wiggler and the \EM\ wave have the  same frequency/wave number, but propagate in the opposite direction.} For example, for $\gamma \sim 10^3$,  $r\sim R_{NS} \sim 10^6$ cm (the radius of a neutron star) and the wave length of the \EM\ waves in lab frame $ \lambda_{EM,b}  \sim$ one centimeter, 
 \be
 \lambda_{EM,b}   \gamma \sim R_{NS}/\gamma \sim 10^3 \, {\rm cm}
 \ee

Thus,  the wiggler (the \Alfven wave) is seen by the particle  as an-EM waves. It is then  Compton scattered to 
\be
\om = 4 \gamma_b^2 ( c k_{w})
\label{om} 
\ee
(The  \EM\ wigglers are similar to static wigglers (with the exception that the resonant frequency is $4 \gamma_b ^2 ( c k_{w})$ for \EM\ wigglers  as opposed to $2 \gamma_b ^2 ( c k_{w})$ for static ones).

 In the frame of the beam the wiggler and the \EM\ wave propagate in opposite direction. Addition of two counter-propagating waves creates a standing wave in the  beam frame. The radiation energy density is smaller at the nodes of the stating wave: this creates a ponderomotive force that pushes the particles towards the nodes -  bunches are created.  
 These bunches are still shaken by the \EM\ wiggler: they emit in phase, coherently.

The above-described model is a variation of a  free electron laser (FEL) mechanism  \citep{Ginzburg47,1951JAP....22..527M,1971JAP....42.1906M,1975SvPhU..18...79Z,1976PhLA...59..187C,1977PhRvL..38..892D,1989SvPhU..32..200A,1989PhFlB...1....3R,1991RvMP...63..949C,freund1986principles}. FELs are operational laboratory devices that   have  high efficiency of energy transfer from the kinetic energy of particles  to the coherent radiation. The FELs in   \mss\ of pulsars and magnetars  will   operate in the somewhat unusual  regime  of ultra-strong guide field, when the cyclotron frequency  $\om_B =  e B_0/(m_e c) $ ($B_0$ is the guide field)  is much larger than the plasma frequency $\om_{p,b}$  and the radiation frequency  $\om$,   $\om_B \gg \om_{p,b}, \om$ \citep{1974PhFl...17..463M,1979PhFl...22.1089K,1980PhFl...23.2376F,2013PhRvS..16i0701G}. 
Astrophysical \Alfven masers were considered by \cite{AlfvenMaser}.
Strongly guide-field dominated regime of FEL, in a relativistic regime of $v_A \sim c$, remains mostly unexplored 
\citep[nonlinear effects in this regime were considered by][]{1976JPlPh..15..335K,1986PhR...138....1S,2006RvMP...78..591M,2020PhRvE.102a3211L}.
 As we demonstrate in \S \ref{Guidefielddominance}, in the \NS\ setting the wiggler frequency in the beam frame is always smaller than the cyclotron frequency. Thus, no Landau level transitions are excited: FEL operated in the guide field-dominated regime. But taking account of  non-resonant drift motions is  important.
Operation of FELs in guide-field dominated regime is the major  part of the present work.

A particularly relevant FEL regime is the SASE process - Self-amplified spontaneous emission. Spontaneous emission is first produced due to incoherent  single particle emission in a wiggler field. Then the  beat between the this   emission and the wiggler  leads to  the parametric resonance, whereby under certain conditions the \EM\ field is further amplified. The SASE process has the right ingredients for the astrophysical applications, when no engineer can tune the parameters of the beam and of the wiggler.

\section{Nonlinear magnetic Thomson scattering }

We start by considering nonlinear  Thomson scattering in a guide-field dominated regime, and apply to model to explain emission bands observed in Crab pulsar and in Fast Radio Bursts. In the  relevant regime the particles mostly experience E-cross-B drift, so that emitted polarization is approximately orthogonal if compared with unmagnetized case. For circularly polarized \Alfven wave, only the fundamental is emitted, while linearly polarized wave produces a set of nearly equidistant 
 emission stripes, resembling those 
 observed in the High Frequency Interpulse of the Crab pulsar.
 
 \subsection{Guiding  field dominance}
 \label{Guidefielddominance}
 The motions of particles in \NS's \mss\ proceeds almost exclusively along the magnetic field, due to high radiative losses.  Interaction with the  wiggled \Bf\  may, in principle, excite gyration motion (in a sense of  transitions between Landau levels).   This  does not happen, as we demonstrate next.

 Condition for resonant excitation by the wiggler
 \ba &&
\om_b= 2   \gamma_ b (k_{w}c) = \om_B
\nn &&
\om_b = \frac{\om}{2 \gamma_b}
   \ea
where $\om_b$ is the wave frequency in the beam frame.
 Together with (\ref{om}) give near the surface of a \NS\ with magnetar's nearly quantum critical \Bf\ 
 \ba &&
 \frac{\om_b}{\om_B} \approx  \frac{1}{ b_q \gamma_b} \frac{\om \lambda_C}{c} = 4 \times 10^{-15} b_q^{-1} \nu_9 \gamma_{b,3} \ll 1
 \nn &&
b_q = \frac{B_{NS}}{B_q}
\nn  &&
B_q= \frac{ c^3 m_e^2}{e \hbar }
\nn &&
 \lambda_C = \frac{ \hbar }{m_e c}
 \ea
In the case of Crab pulsar, near the \LC, the corresponding value is still small, $\sim 10^{-7}$, though  could be larger in regions of low \Bf, in the current sheet. 

 We are interested in the specific new regime of FEL in the \mss\ of \NSs,  highly guide field-dominated regime. We neglect possibility of cyclotron excitation

 \subsection{Magnetic nonlinearity parameters $a_H$ and  $a_{H,b}$} 

 It is assumed that \Alfven waves are of low amplitude,  in a sense  that the fluctuating \Bf\ $ B_w$  (measured in the lan frame)
 is much smaller than the guide field $B_0$, 
 \be
 a_{H}  = \frac{B_w }{B_0} \ll 1
 \label{aH1}
 \ee
  
 Still, such ``weak''  waves   will   have exceptionally high intensities, much larger than anything that could be expected in high frequency EM waves. 
For example,  \Alfven wave in critical quantum field  with the wavelength of \NS\ radius will have  laser intensity parameter  \citep{1975OISNP...1.....A}
\ba &&
a_A \equiv \frac{e B_w}{m_e c \om}  = \frac{a_{H}b_q}{2 \pi} \frac{R_{NS}}{\lambda_C} = 4  \times 10^{15} a_{H}  b_q  \gg 1
\label{aa}
\label{aH}
\label{bq}
\ea
\Alfven waves in high magnetized plasma propagate with relativistic velocities, hence $E_w \sim B_w$. Parameter $a_A$ is Lorentz-invariant.

 Thus, even weak  \Alfven waves, with relative amplitude  $10^{-16} \ll a_{H} \ll 1$, are in fact of extremely  high intensity by the laser standards.  Scattering of these high intensity waves then converts beam energy into  high frequency  escaping radio waves.  The scattering is done by  a beam of high energy particles generated during reconnection events.
 
 The laser intensity parameter (\ref{aa}) characterizes interaction of unmagnetized particle with an EM wave. 
 Interaction of a strongly magnetized particle with the regular EM field, is characterized by magnetic non-linearly parameter  $a_{H}  \ll 1$, Eq. (\ref{aH1}), with a change of wiggler $B_w$ to EM wave amplitude
 \citep[Eq. (\ref{aH}));][]{2016MNRAS.462..941L,2019arXiv190103260L,2020PhRvE.102a3211L}. Qualitatively, motion of a particle in a wave in this case  is a relatively slow E-cross-B drift.

   Magnetic  intensity parameter  $a_{H} $ is {\it not} Lorentz-invariant: if a particle is moving with \Lf\ $\gamma$, then in it's rest frame
   \be
   a_{H,b} = \gamma_b a_{H}  
   \ee
   (This follows from the Lorentz transformation of the field component perpendicular to the velocity.)
   This is a highly important effect: even for very low amplitude perturbations, $a_{H}  \ll 1$, the effects of the \EM\ wave (the wiggler) are amplified in the particle rest frame by a factor $\gamma_b \gg 1$. 
   

\subsection{Particle trajectories in EM pulse with guiding \Bf}
Generally, nonlinear effects cannot be treated  as basic Fourier analysis, hence we first  derive  general relations  for arbitrary profile of an \EM\ pulse.


On basic grounds the \EM\ fields in the particle frame  are 
\ba &&
\B= \{a_{H,b} f_w' , a_{H,b} g_w',1\} B_0 
\nn &&
\E= \{-  g_w' , f_w',0\} a_{H,b}  B_0
\nn &&
{\xi_+}= k_{w,b}( t+z )
\ea
where $f_w \equiv f_w(\xi_+)$ and $g_w \equiv g_w(\xi_+)$ are corresponding vector potentials and speed of light is set to unity.


Equations of motion,
\ba &&
\dot{\beta_x} = \left( \beta_y -  a_{H,b} g_ w' (1+\beta_z)  \right)\om_B
\nn &&
\dot{\beta_y} = \left(- \beta_x +  a_{H,b} f_ w' (1+\beta_z) \right)\om_B
\nn &&
\dot{\beta_z} = ( g_w' \beta_x-f_w' \beta_y)a_{H,b} \om_B ,
\nn &&
\om_B = \frac{ e B_0}{m_e c}
\label{eq11} 
\ea
for non-relativistic motion in the frame of the particle,  $(1+\beta_z)  \approx 1$, the set (\ref{eq11}) becomes
\ba &&
\dot{\beta_x} = \left( \beta_y -  a_{H,b} g_ w'   \right)\om_B
\nn &&
\dot{\beta_y} = \left(- \beta_x +  a_{H,b} f_ w ' \right)\om_B
\nn &&
\dot{\beta_z} =( g_w' \beta_x-f_w' \beta_y)  a_{H,b} \om_B
\ea

In the magnetically-dominated  limit $\om_B \rightarrow \infty$,
\ba &&
\beta_x \approx  a_{H,b} ( f_w - \alpha g_w')
\nn &&
\beta_y \approx a_{H,b} (g_w+ \alpha F _w)
\nn &&
\alpha = \frac{ k_{w,b}}{\om_B}
\nn &&
\dot{\beta_z}  = 
 -  \partial_z \left(  f_w^2 + g_w^2\right)   \frac{a_{H,b}^{2}}{2}  
 \label{beta}
 \ea
 where in relation for $\dot{\beta_z}$ we neglected terms $\propto  a_{H,b}^{2}/\om_B \propto 1/\om_B^3$.

With appropriate switch-on conditions,
 \be
 \beta_z = -\left(  f_w^2 + g_w^2\right) \frac{ {a_{H,b}^{2}  } }{2} 
\ee

For harmonic wiggler
\ba &&
f_w = \sin \xi_+
\nn &&
g_w= - \eta_g \cos  \xi_+, 
\ea
where parameter  $\eta_g $ controls polarization,  $\eta_g = 0$ for linear and  $\eta_g = \pm  1 $ for circular (to avoid unnecessary  complications we do not normalize to constant wave intensity).

 In the limit of small axial oscillations, $(\Delta z) k_{w,b}\ll 1$, we find
 \ba &&
 {\bf r}= 
  \{ (1 - \alpha \eta_g )  \sin {\xi_+} \left( \frac{a_{H,b}}{k_{w,b}} \right)   ,    (\alpha- \eta_g)   \cos {\xi_+}\left( \frac{a_{H,b}}{k_{w,b}} \right) ,   v_z t -(\Delta z) \sin(2 k_{w,b}t)\}
  \nn && 
   \v= 
  \{ (1 - \alpha \eta_g )  \cos {\xi_+}a_{H,b} ,    (\eta_g - \alpha )   \sin {\xi_+}a_{H,b},  v_z -(\Delta v) \cos(2 k_{w,b}t )\}
 \nn &&
   {\bf a} =
  \{ - (1 - \alpha \eta_g )  \sin {\xi_+} {a_{H,b}}{k_{w,b}}   ,    (\eta_g - \alpha )   \cos {\xi_+}{a_{H,b}}{k_{w,b}} ,  4 k_{w,b}^2 (\Delta z) \sin(2 k_{w,b}t)\}
  \nn &&
  v_z = - (1+\eta_g^2) \frac{ {a_{H,b}^{2}  } }{4}
 \nn  &&
 (\Delta v)= (1-\eta_g^2) \frac{ {a_{H,b}^{2}  } }{4}
  \nn  && 
  (\Delta z) =(1-\eta_g^2) \frac{ {a_{H,b}^{2}  } }{8 k_{w,b}}
 \ea

For linearly polarized wiggler, $\eta_g =0$, the particle trajectories.
\ba && 
k_{w,b}^2 x^4 - a_{H,b}^{2} x^2 + 16 z^2=0
\nn && 
k_{w,b}^2 x^4 - a_{H,b}^{2}  \alpha^2 x^2 + 16 \alpha^4 z^2=0
\ea 
These are figure-8 trajectories both in the $x-z$ and $y-z$ planes, Fig. \ref{X-Y-Z-3D}.

\subsection{Numerical examples: single particle  motion }
\label{Simulations1}

We developed two  codes (in {\it Mathematica} and {\it Python})  to trace the motion of a particle in EM field based on Boris integrator \citep{birdsall}. Results of numerical stimulations were cross-checked with  the two codes.

\begin{figure}[h!]
\centering
\includegraphics[width=.99\textwidth]{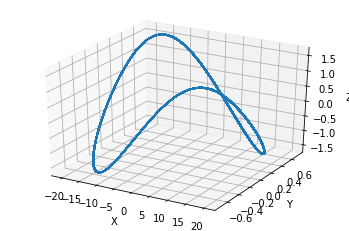} 
\caption{3D rendering of particles trajectories in the beam frame.}
\label{X-Y-Z-3D} 
\end{figure}

\begin{figure}[h!]
\centering
\includegraphics[width=.32\textwidth]{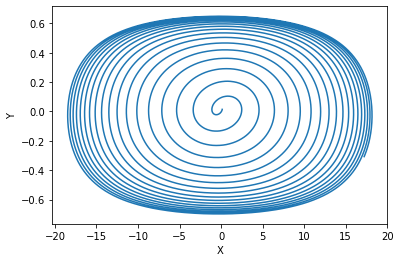}
\includegraphics[width=.32\textwidth]{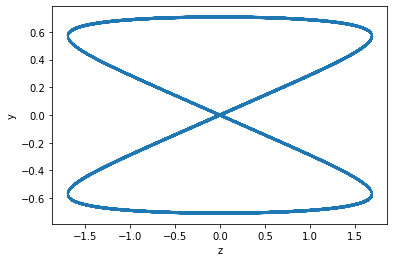}
\includegraphics[width=.32\textwidth]{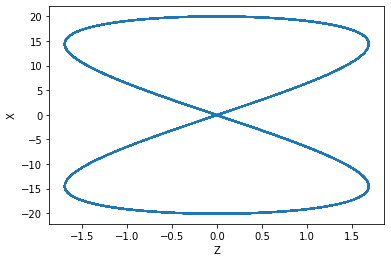}  
\caption{Closer look at particle trajectories in the case of  linearly polarized EM wiggler with dominant guide field. The Y-X plane (left panel) illustrates effects of adiabatic switching of the wave: steady state trajectory (outer envelope) is reached on few times for adiabatic switching. The X-Z and Y-Z cuts show trajectories after  the   steady state is reached. Thinness of the curves  composed of multiple tracks  illustrate the stability of the codes, and  of the adiabatic switching procedure.   Largest amplitudes of oscillations are in the wiggler plane $x-z$.   Amplitudes of oscillations are not normalized between Fig. \protect\ref{X-Y-Z-3D}}
\label{Yegor-single} 
\end{figure}

In Fig.  \ref{X-Y-Z-3D} -\ref{Yegor-single} we plot particle trajectories of particles subject to the \EM\ wave in a guide field dominated regime, Fig. \ref{Yegor-single}. To take correct account of particle's dynamics we implemented adiabatic switching of the wave intensity.
Particle makes a saddle-like motion, Fig. \ref{X-Y-Z-3D}

\begin{figure}[h!]
\centering
\includegraphics[width=.32\textwidth]{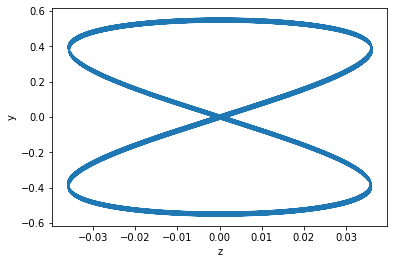} 
\includegraphics[width=.32\textwidth]{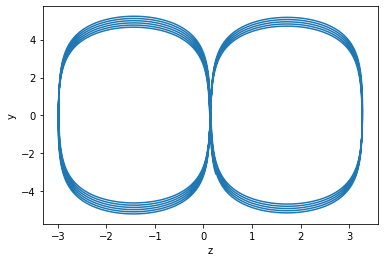}
\includegraphics[width=.32\textwidth]{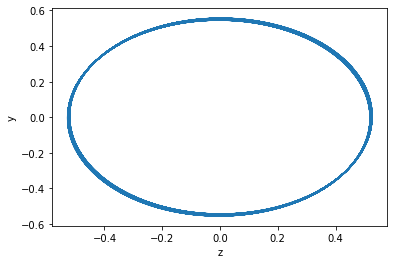}
\caption{Particles trajectories for strong static wiggler case (no guide field, beam frame). Fixed particle energy, wiggler strength increases left to right.  At sufficiently high field (right panel) a particle just executes elliptical  Larmor motion within a given field patch.  The intermediate case is a curious one:  a particle experiences a Spicer-like orbit mostly along the  $y$ axis ({\it not} along $z$ axis as in the limit of small wiggler field or EM wave, left panel). The strong wiggler  regimes  (horizontal figure 8 and the ellipse) are not applicable to pulsar \mss. }
\label{FIG8} 
\end{figure}

(In passing we note that for  static wiggler  the particle trajectory {\it does} depend on the wiggler strength, changing from ``vertical 8" to ``horizontal 8", to ``circle'': for sufficiently strong  wiggler a particle just executes Larmor gyrations, Fig. \ref{FIG8}. The latter regimes are not relevant for pulsars.)

\subsection{Single particle emissivity} 
\subsubsection{Linear regime}

Instantaneous and average powers  (neglecting terms of the order of $a_{H,b}^4$) are
\ba &&
W = \frac{2}{3} \left(  (\alpha -\eta_g)^2 \cos^ {\xi_+}+(1-\alpha \eta_g)^2 \sin^2 {\xi_+}\right)  \frac{a_{H,b}^{2} e^2 k_{w,b}^2}{c^3}  
\approx \left(  1- 4 \alpha \eta_g + \eta_g^2 -(1-\eta_g^2)   \right)    \frac{a_{H,b}^{2} e^2 k_{w,b}^2}{3c^3} 
\nn &&
W_a =  \left(  1- 4 \alpha \eta_g + \eta_g^2  + \alpha^2 (1+\eta_g^2) \right)    \frac{a_{H,b}^{2} e^2 k_{w,b}^2}{c^3} 
\approx \left(  1- 4 \alpha \eta_g + \eta_g^2   \right)    \frac{a_{H,b}^{2} e^2 k_{w,b}^2}{3c^3} 
\ea
where the last relations neglect $\alpha^2$.

Energy flux in the wave
\be
S_w = (1+\eta_g^2 ) \frac{  B_0 ^2}{8\pi} a_{H,b}^{2} 
\ee
The cross-section
\ba &&
\sigma = \frac{W_a}{S_w} = \frac{8\pi}{3}\left(  \frac{1- 4 \alpha \eta_g + \eta_g^2}{1+\eta_g^2}   \right)   \frac{ (k_{w,b}c)^2}{\om_B^2}  \frac{e^4}{m_e^2 c^4}=
\left(  \frac{1- 4 \alpha \eta_g + \eta_g^2}{1+\eta_g^2}   \right)  \frac{ (k_{w,b}c)^2}{\om_B^2}     \sigma_T=
\nn &&
 \frac{ (k_{w,b}c)^2}{\om_B^2}     \sigma_T \times
 \left\{
 \begin{array}{cc}
 1, &  \mbox{for linearly  polarized $\eta_g=0$}
 \\
 1\pm 2 \alpha, & \,  \mbox{for circularly   polarized $\eta_g=\pm 1$}
 \end{array}
 \right.
\ea
Two signs for the circularly polarized wave correspond to two combinations of the wave polarization and direction of rotation of a particle.

For linearly polarized wave, $\eta_g =0$ 
instantaneous acceleration is 
\be
{\bf a} = \{ - \sin \xi_+,  \alpha \cos \xi_+,0\} a_{H,b} k_{w,b}c^2
\ee
thus, scattered waves are elliptically polarized. In the limit $\alpha \to 0$
the differential cross-section
\ba && 
\frac{d \sigma}{d \Omega} = \frac{3}{8\pi} \sigma  \sin^2\Theta
\nn &&
\cos \Theta  = - \cos \phi \sin \theta
\ea
where $\theta, \, \phi$ are the direction of photon propagation.
For unpolarized lite
\be
\frac{d \sigma}{d \Omega} =  \frac{3}{16\pi} \sigma (1+\cos^2 \theta)
\ee

\subsection{Non-linear effects: multiple bands}
 We are interested in the back-scattered X-mode, propagating along the field lines. 
 Using the emission formula
 \ba && 
 \frac{d^2 I}{d \om d\Omega} = \frac{ e^2 \om^2}{4\pi^2 c} [  {\bf n}  \times ( {\bf n}  \times  \tilde{ \v}) ]^2
 \nn &&
 \tilde{ \v} =\int dt \v e^{-i \phi} 
 \nn &&
\phi= \om (t-  {\bf n} \cdot{\bf r} )
\nn &&
{\bf n} =\{0,0,1\}
 \nn && 
 \tilde{ \v} =\int dt
  \tilde{ \v}_0 e^{-i \phi} 
  \nn &&
  \tilde{ \v}_0 =  \{ - (1-\alpha \eta_g)   ,  (\eta_g - \alpha f_w')   , 0\}  \frac{a_{H,b} }{2} 
  \nn &&
  \phi= (\om-k_{w,b}) t + (\Delta z) (\om+ k_{w,b}) \sin 2 k_{w,b}t
  \label{phase1}
  \ea
  
  To take account of the jitter oscillation $ (\Delta z)  \propto a_{H,b}^{2}$, we  use Jacobi-Anger expansion
of Bessel functions
\be
e^{i b \sin \sigma} = \sum_n J_n(b) e^{i n \sigma}.
\label{JacobiAnger}
\ee
We find
\be
  \tilde{ \v} = \sum_n  \tilde{ \v}_0  J_n \left(  (\Delta z) (\om+k_{w,b})   \right)  \int dt  e^{i t( \om - (1+2n) k_{w,b})}
  \ee
  Thus, the resonant condition is
  \be
  \om = k_{w,b} (1+2 n)
  \label{kb}
  \ee
  A particle produces a set of spontaneous emission bands separated in frequency by $2 n k_{w,b}$
  
  The strength of each resonance (Fourier power of velocity)   scale as
\be
\propto  J_n \left( \frac{1-\eta_g^2}{4} (1+ n)  a_{H,b}^{2} \right) ^2 
\ee
Circularly polarized wave, $\eta_g=1$ produces only the fundamental mode ($n=0$, $\om=k_{w,b}$)  with
scattering cross-section
\ba &&
\frac{d^2 I}{d \om d\Omega} =(1-\alpha)^2\frac{e^2 \om^2 }{4  \pi} \delta(\om - k_{w,b})
\nn &&
\sigma = (1-\alpha)^2 \left( \frac{  c k_{w,b}} {\om_B} \right) ^2 \left( \frac{  e^2 } {m_e c} \right) ^2
\ea

For linearly polarized incoming wave, $\eta_g=0$
the strength of each resonance scales as 
\be
\propto  J_n \left( \frac{1}{4} (1+ n)  a_{H,b}^{2} \right) ^2 \approx \frac{ (n)^n} { 2^{2n+1} n!} a_{H,b}^{2 n}
\ee
The strength of the resonance  is  determined  by the order of the resonance $ n$ (for higher $n$ the argument of the Bessel function is larger; this partially compensates the fact that for high $n$ the values of Bessel functions near zero are smaller).  Thus, even for small fluctuation of the field $a_{H,b}\ll 1$,  higher order resonances may, under certain conditions, become comparable to the basic one. 

\begin{figure}[h!]
\centering
\includegraphics[width=.99\textwidth]{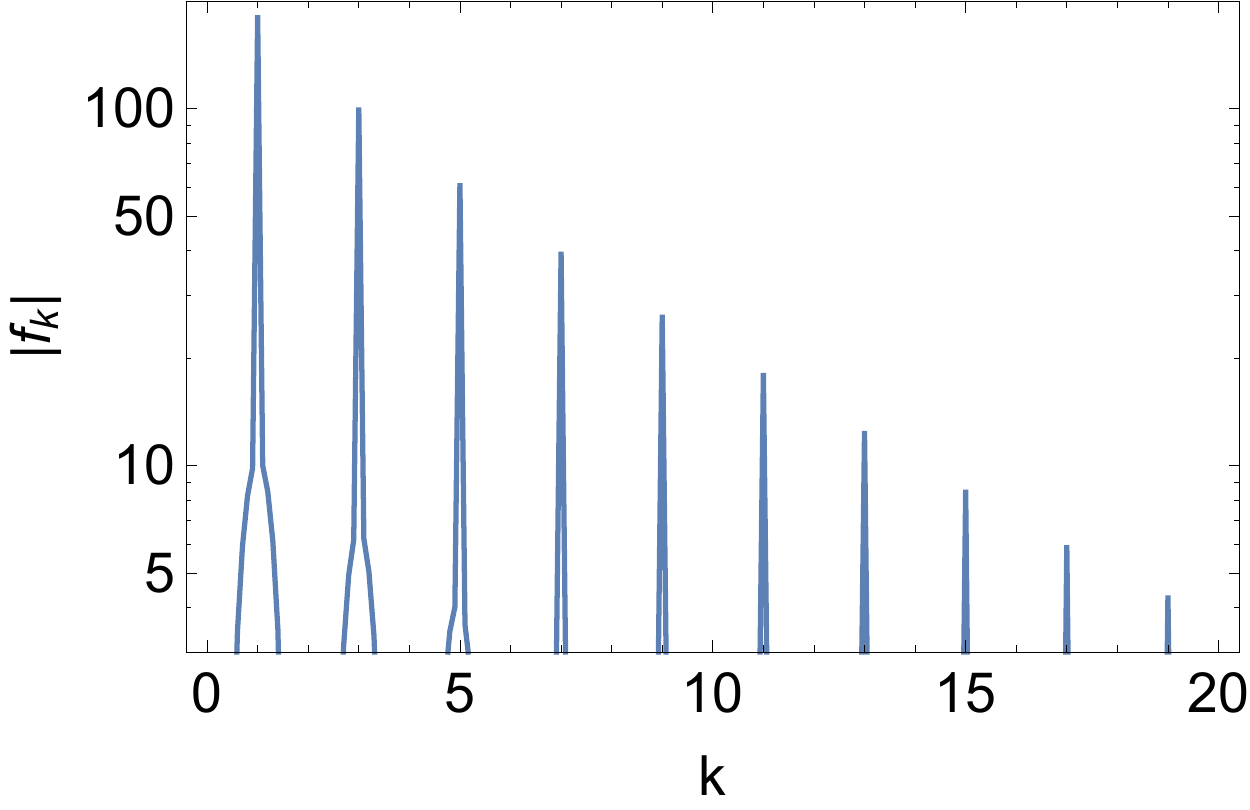} 
\caption{Numerical Fourier transform of particle's current (\protect\ref{phase1}) with the jitter amplitude $ (\Delta z)=0.3$ and $k_{w,b}=1$. Clear resonances are seen at $\om/k_{w,b}= 1+  2n$ (this is the direct  numerical demonstration of the  Jacobi-Anger expansion (\protect\ref{JacobiAnger})}
\label{stripes} 
\end{figure}

As a check we performed 
numerical integration of (\ref{phase1}); this,  naturally, gives the Jacobi-Anger expansion/resonances, Fig. \ref{stripes}

\subsection{Astrophysical applications of nonlinear scattering: Emission band in Crab's spectrum} 

One of the most interesting  property of Crab  radio emission is observations of numerous emission stripes in Crab High Frequency interpulse 
\citep{1999ApJ...522.1046M,eilek07,2016ApJ...833...47H,2016JPlPh..82c6302E}.
Fig. \ref{Fig4Right}.
Emission is composed of at least a dozen relatively  narrow emission bands, almost equally spaced in frequency
(spacing increases with frequency at a rate of $6\%$).
 We consider  these stripes being the properties of the emission mechanism, not propagation \citep{2007MNRAS.381.1190L}. It is  one of the most intricate properties of pulsar radio emission - explaining it will be a strong argument in favor of any model of pulsar radio emission.

There is a caveat regarding the band separation in the Crab's HFIP: the observed stripes in Crab are slightly non-equidistant, band separation increases with observing frequency at a rate of 6\%
\citep{2016ApJ...833...47H,2016JPlPh..82c6302E}. We consider this a next-order effect. Most importantly,  equal spacing (\ref{kb}) is for single particle emission - non-linear saturation effects may slightly modify the peak frequencies. 
As a possible route to explain this slight upward shift in frequency separation observed in Crab \citep{2016ApJ...833...47H,2016JPlPh..82c6302E},  in comparison with regular spacing (\ref{kb}),  we offer the following argument. Qualitatively, harmonic wiggler creates periodic variations of the refractive index. The propagation of waves is then described by Hill's and/or Matthieu's equation (a parametrically-driven oscillator). For small amplitudes of driving the  periodic solutions are separated by a constant  frequency  difference.  At the weakly non-linear stage,  the ``Floquet tongues'' of the Mathieu's equation (regions of periodic solutions) generally move {\it slightly to higher frequencies}. For example, for  Mathieu equation written in the form
$\ddot{x} + ( \om^2 +\epsilon \cos( 2 t) ) x=0$, $\epsilon \ll 1$, then   the  center of the  second band is located at $< \om_{2} >=2  (1+ \epsilon^2/192)$ \citep{1969JMP....10..891C,Verhulst2009}. The shift to slightly higher frequency is a general property for harmonic  band's number larger than unity. Thus, nonlinear effects at saturation are expected to shift the band to slightly {\it highly} frequencies, as observed. (Beside nonlinearity additional effect that affects the emitted frequency is angle of propagation of the emitted radiation with respect to the structure in the dielectric tensor of the medium, similar to the case of  Bragg's scattering.)

\section{Free Electron Laser in guide field dominated regime}
 
 \subsection{Particle motion in guide-field dominated EM wave and the wiggler}
  
Consider orthogonally polarized harmonic  EM  modes:  wiggler mode  with amplitude $a_{H,b}$ (propagating ``to the left'') and EM mode with amplitude $a_{EM,b}$ (propagating ``to the right'').
In the beam frame both waves have the same frequency/wave number $k_{w,b}$.
Using parametrization of the vector potential 
\ba &&
\A = \left( a_{H,b}  \{\eta_g \sin \xi_+, \cos \xi_+,0\} +
a_{EM,b} \{\cos \xi_-, \eta_{EM,b} \sin \xi_-,0\}
\right)  \frac{B_0}{k_{w,b}} G(t) 
\nn &&
\B = \curl \A+ B_0 {\bf e} _z
\nn &&
\E = - \partial _t \A
\nn && 
\xi_- =  k_{w,b} (t-z)
\nn &&
\xi_+ =  k_{w,b} (t+z)
\label{AAA}
\ea
where $G(t)$ is some switch-on function. Parametrization (\ref{AAA}) carries non-zero current components $j_{x,y} \propto \dot{G}, \ddot{G}$.
Both the adiabatic switch-on terms, and  the $\propto 1/\om_B$ terms are needed to get the correct phasing and force-balance.

Average Poynting flux is
\be
F_z =\left((1+\eta_{EM,b}^2) a_{EM,b}^2 - (1+\eta_{g}^2) a_{H,b}^2 \right) \frac{B_0^2}{8\pi} 
\ee

Assuming non-relativistic motion in the beam frame (so that $1+ \beta_z \to 1$), neglecting effects of charge separation (this would correspond to the Compton regime of FEL),  equations of motion are (switch-on terms are omitted for clarity;  full system of equations is used for numerical calculations)
\ba &&
\dot{\beta}_x =  \left(a_{{EM,b}} \sin \left(\xi _-\right)-a_{H,b} \eta _g \cos \left(\xi
   _+\right)+\beta _y\right)  \omega _B
   \nn &&
   \dot{\beta}_y=  B \left(-a_{{EM,b}} \eta _{{EM,b}} \cos
   \left(\xi _-\right)+a_{H,b} \sin \left(\xi _+\right)-\beta _x\right)\omega _B
   \nn &&
   \dot{\beta}_z =
   \left(a_{{EM,b}} \left(\sin \left(\xi _-\right) \beta _x-\eta _{{EM,b}} \cos
   \left(\xi _-\right) \beta _y\right)+a_{H,b} \left(\eta _g \cos \left(\xi _+\right) \beta
   _x-\sin \left(\xi _+\right) \beta _y\right)\right)\omega _B
         \ea

 Leading terms in $1/\B_0$ expansion give 
\ba && 
\beta_x =
\left(\alpha -\eta _{{EM,b}}\right) a_{{EM,b}} \cos \left(\xi _-\right) +  \left(1+ \alpha  \eta _g\right) a_{H,b}
   \sin \left(\xi _+\right)
\nn &&
\beta_y =- \left(1-\alpha  \eta _{{EM,b}}\right)a_{{EM,b}} \sin \left(\xi
   _-\right) +a_{H,b}   \left(\alpha +\eta _g\right)\cos \left(\xi _+\right)
\nn &&
\dot{\beta}_z= \delta  \left(\left(1-\eta _{{EM,b}} \eta _g\right) \cos \left(2 z
   k_{w,b}\right)-\left(1+\eta _{{EM,b}} \eta _g\right) \cos \left(2 t k_{w,b}\right)\right)
  \label{1}
\\ &&
\delta=   a_{H,b} a_{EM,b}   \om_B
\label{01}
\ea
Notice that for   $z$-oscillations  the leading terms in parameter $\delta$ is  ${\cal{O}}(1/B_0) $ (and not ${\cal{O}}(1/B_0^2)$).

The energy of a particle $\epsilon$ evolve according to 
\be
\partial_t \epsilon= -a_{{EM,b}} a_{H,b} k_{w,b}\left(\eta _{{EM,b}}+\eta _g\right) \cos
   \left(2 t k_{w,b}\right)+\frac{1}{2} a_{{EM,b}}^2 \left(1-\eta _{{EM,b}}^2\right)
   k_{w,b}\sin \left(2 \xi _-\right)+\frac{1}{2} a_{H,b}^2 \left(1-\eta _g^2\right) k_{w,b}\sin
   \left(2 \xi _+\right) \propto B_0^{-2}
   \ee
   
The system (\ref{1}) is the main set of equations governing particle motion in the combined fields of the wiggler and the EM wave.
Qualitatively, the transverse motion is dominated by the electric drift, $\beta_\perp \approx \E\times \B_0/B_0^2$.

\subsection{Dynamics of trapped particles: ponderomotive Hamiltonian}

Longitudinal trajectories, Eq. (\ref{1}),  consist of fast jitter  $ \propto \cos ( 2 k_{w,b}t)$  and slow oscillations  $ \propto \cos ( 2 k_{w,b}z)$.
We can rewrite  Eq. (\ref{1}) as
\be
\partial_t \left( \frac{\beta_z^2}{2}  + \frac{\delta}{2 k_{w,b}}  (1-\eta_{EM,b} \eta_g)  \sin ( 2  k_{w,b} z) \right) =  -   \beta_z  (1+\eta_{EM,b} \eta_g) \delta  \cos ( 2  k_{w,b} t),
\ee

The slow oscillations are governed by 
\ba &&
\partial_t \beta_z = \tilde{\delta} \cos ( 2 k_{w,b}z)
\nn &&
\tilde{\delta} =  (1-\eta_{EM,b} \eta_g) \delta
\label{33}
\ea
(Note that only  for circularly polarized EM and wiggler  $\eta_{EM,b} =\eta_g=1$ the slow oscillations vanish.)
Below we drop the tilde sign over $\tilde{\delta}$.

Relations (\ref{33})  allow a first integral
\ba &&
\frac{\beta_z^2}{2} + V_p=  {\rm Const} 
\nn && 
V_p = \frac{\delta}{2 k_{w,b}} \sin (  2 k_{w,b}z) 
\ea
where $V_p$ is the ponderomotive potential.

Next we shift the coordinates so that the O-point in $\beta_z -z$ plane is at $z=0$ $z \to z - \pi/(4 k_{w,b}) $, and parametrize  the ${\rm Const}$ by the value of the velocity $\beta_z =\beta_0$  at $z=0$:
\be
V_p = \frac{\delta}{ k_{w,b}}  \sin ^2(  k_{w,b}z)
\label{Vp}
\ee
This is the ponderomotive potential.
The O-point has $\beta_z= \beta_0=0$ and is located at $z=0$.

The Hamiltonian then takes the form
\ba &&
{\cal H} = \frac{\beta_z^2}{2} + V_p =  \frac{\beta_0^2}{2}
\nn &&
\partial_t z = \partial_{\beta_z} {\cal H}  = \beta_z
\nn &&
\partial_t \beta_z= - \partial_{z} {\cal H}=- \delta \sin (2 k_{w,b}z)
\label{H}
\ea
(this is different from (\ref{33}) due to the $\pi/4$ phase shift).
Parameter  $\beta_0$ is the maximal velocity by a particles in the combined EM-wiggler field.  

The separatrix is determined by 
\be
\beta_0 =\beta_{S} = \sqrt { 2 \frac{\delta}{ k_{w,b}} }
\label{betaS}
\ee
Normalizing velocities by $\beta_{S} $  the   Hamiltonian becomes
\ba && 
{\cal H} =
\beta_z^2 +   \sin ^2(  k_{w,b}z)= \eta_\beta^2
\nn &&
\beta_z = \beta_z/\beta_{0, S}
\nn &&
\eta_\beta= \frac{ \beta_0}{\beta_{0, S}} \leq 1
\ea

\begin{figure}[h!]
\centering
\includegraphics[width=.99\textwidth]{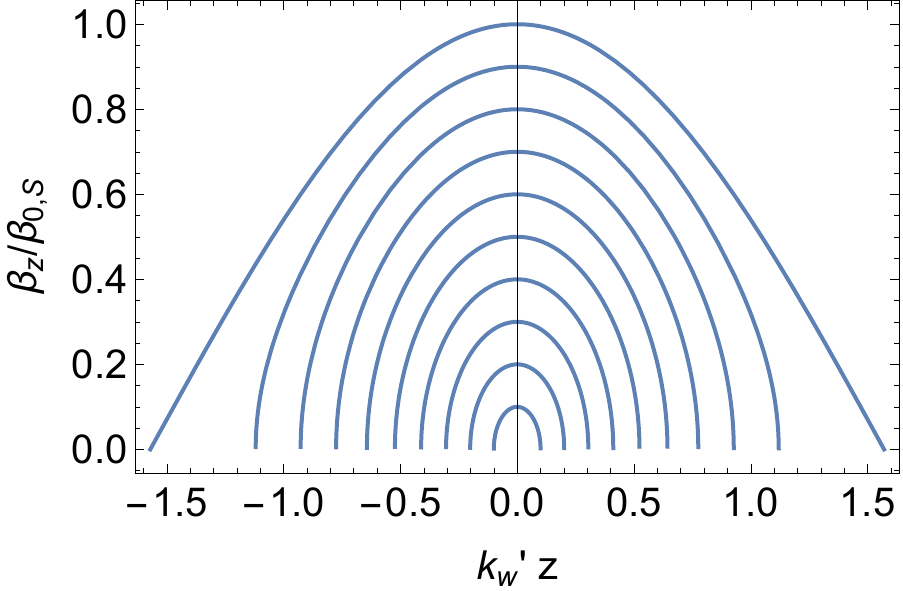} 
\caption{Trajectory of trapped partiles in $\beta_z-z$ plane, Eq.  (\protect\ref{Opoint}), $\eta_\beta = 0.1, \, 0.2...  0.9.$}
\label{Opoint1} 
\end{figure}

The flow lines in the $\beta_z -z$ plane are then given by
\be
{ \beta_z} = \pm
\sqrt{ \eta_\beta^2 - \sin ^2(  k_{w,b}z) },
\label{Opoint}
\ee
Fig. \ref{Opoint1}. For adiabatic switching, all particles are trapped. 

Point where $\beta_z=0$ are determined by 
\be
\sin ( k_{w,b}z) = \eta_\beta
\ee
In coordinates $\beta_z /\beta_S - k_{w,b}z$ the X-point is at $45^\circ$. 

Hamiltonian (\ref{H}) has an adiabatic invariant
\be
{\cal I } =\oint \beta _z dz = \frac{4 \beta_0}{k_{w,b}}  {\cal E }\left(\arcsin \left( \beta_0 \frac{k_{w,b}}{2 \delta} \right), \frac{2 \delta }{k_{w,b}\beta_0^2}   \right) =
\frac{4 \sqrt{2} \sqrt{\delta} }{k_{w,b}^{ 3/2} }\eta_\beta{ \cal E} \left(\arcsin(\eta_\beta), 1/\eta_\beta^2  \right)
\ee
where $ {\cal E }$ is the  elliptic integral of the second kind.

\begin{figure}[h!]
\centering
\includegraphics[width=.32\textwidth]{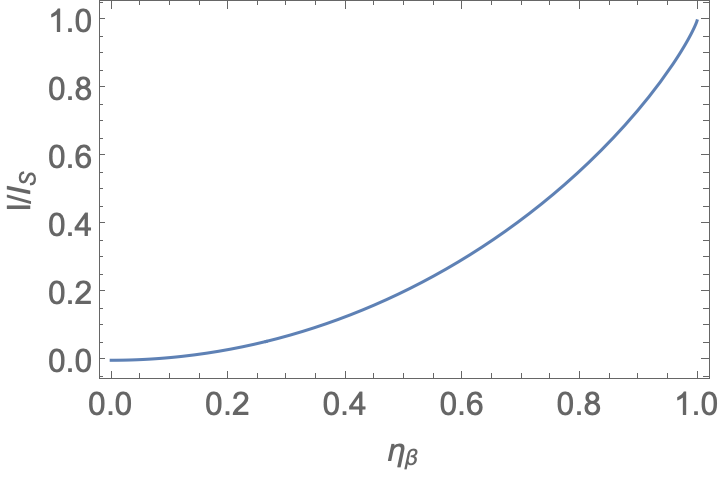} 
\includegraphics[width=.32\textwidth]{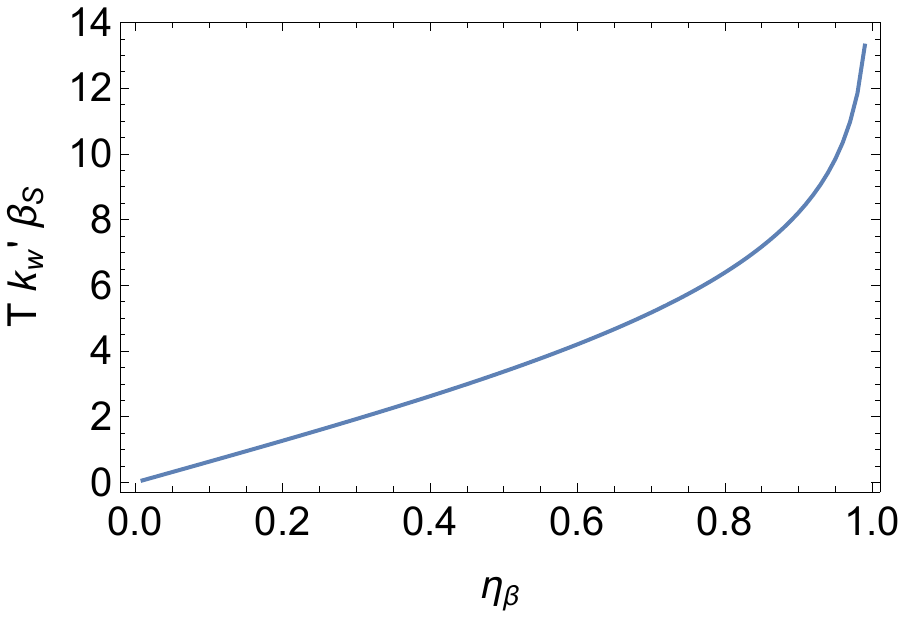} 
\includegraphics[width=.32\textwidth]{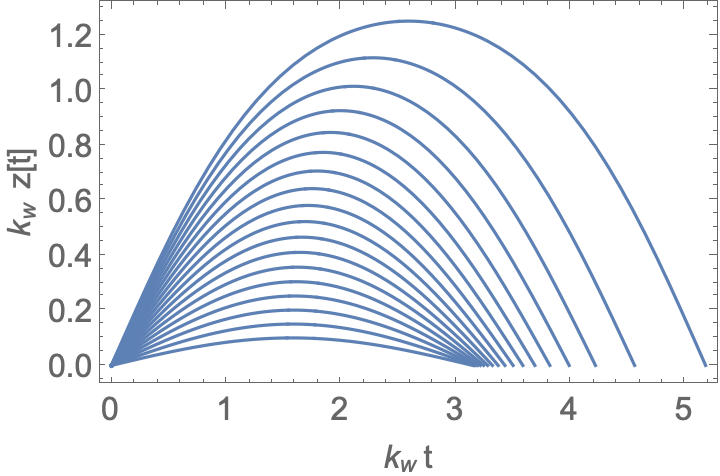} 
\caption{Left Panel: Value of the adiabatic invariant as function of $\eta_\beta$. Central Panel: period of oscillations as function of $\eta_\beta$ (ratio of maximal velocity along the given trajectory to the  maximal velocity on the separatrix). Right Panel: particle trajectories $z(t)$ for different values of $\eta_\beta = 0.05,\, 0.1\, ... 0.95$. }
\label{IAdiab} 
\end{figure}

On the separatrix, $\eta_\beta=1$, 
\be
{\cal I }_S =
\frac{4 \sqrt{2} \sqrt{\delta} }{k_{w,b}^{ 3/2} },
\ee
see Fig. \ref{IAdiab}.


Motion of a particle can be integrated
\ba &&
z k_{w,b}= J_A\left( k_{w,b} \beta_0 t , \frac{1}{\eta_\beta^2} \right)
\nn && 
\beta_z = \beta_0 J_{ND}\left( k_{w,b} \beta_0 t , \frac{1}{\eta_\beta^2} \right)
\label{motion}
\ea
where $J_A$ is Jacobi amplitude and $J_{ND}$ is Jacobi elliptic function. Time is defined so that at time $t=0$ the particle is at $z=0$.

On the separatrix, $\eta_\beta=1$,
\ba &&
z k_{w,b}= Gd(k_{w,b} \beta_0 t )
\nn &&
\beta_z = \beta_{S}\sech(k_{w,b} \beta_S t )
\ea
where $Gd$ is Gudermannian function. Thus, 
 a particle on the separatrix never reaches the point where $\beta_z=0$. 

A period of oscillations can be found as a (four times) the point where $\beta_z$ becomes $0$ in (\ref{motion}), Fig.\ref{IAdiab}.

Slow oscillations are not coherent, in a sense that particles with different parameter $\beta_0$ have random phases - thus, density enhancement is approximately constant in time.

\subsection{Direct numerical integration of the equation of motion}

\begin{figure}[h!]
\centering
\includegraphics[width=.32\textwidth]{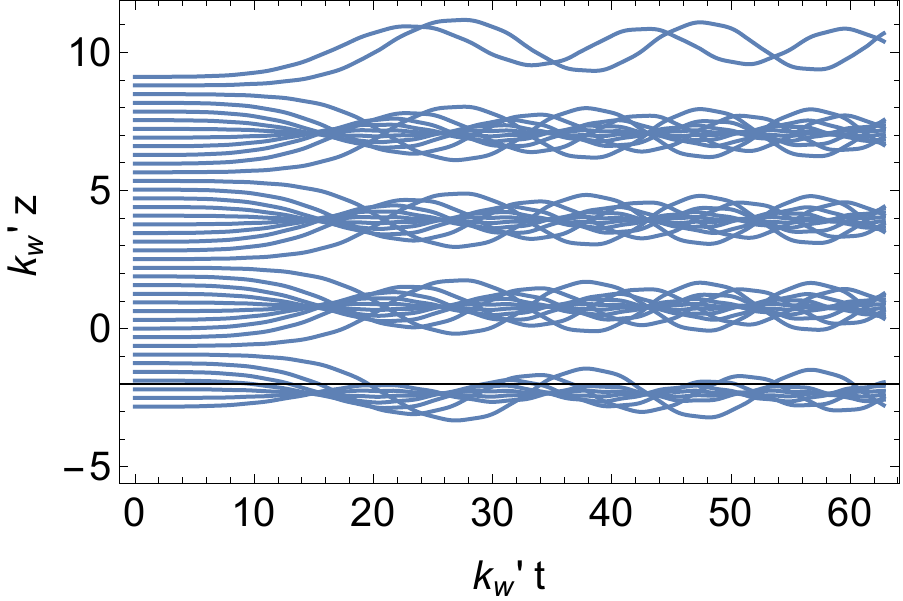} 
\includegraphics[width=.32\textwidth]{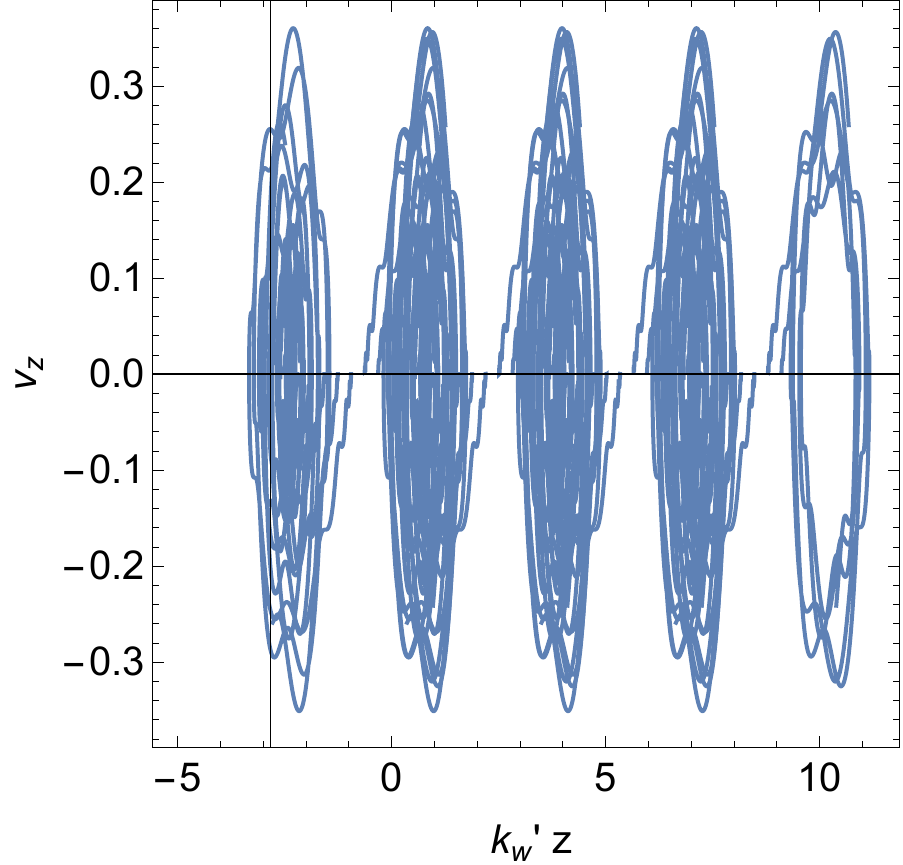}
 \includegraphics[width=.32\textwidth]{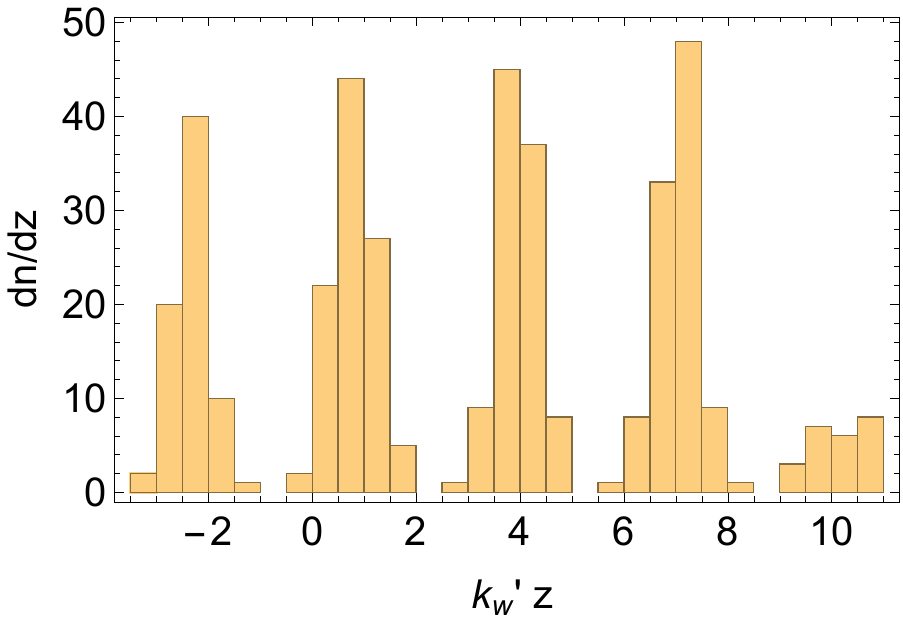}
\caption{ Particle trajectories (left panel), phase diagram (central panel) and distribution (right panel) for particles in counter-propagating wiggler and EM fields
(integration of Eq. (\protect\ref{1}) with initially homogeneous distribution of particles).
In this particular example $k_{w,b}=1$,  the switch-on function is $G=1 - e^{-t/\tau}$ with $\tau = 6\pi$, $\delta = 0.1 $. Particles are initially located at $-\pi < z < 3  \pi$. Two density enhancements per period are clearly seen (they are constant in time at times much longer than the switch-on time. }
\label{trajector} 
\end{figure}

\begin{figure}[h!]
\centering
 \includegraphics[width=.49\textwidth]{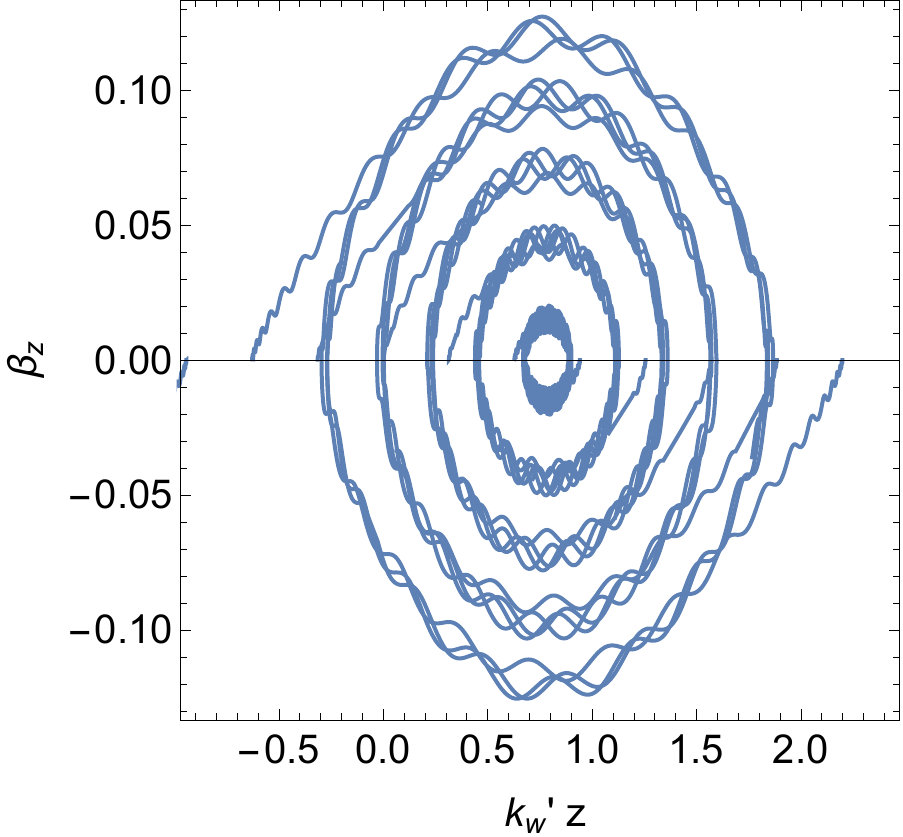}
  \includegraphics[width=.49\textwidth]{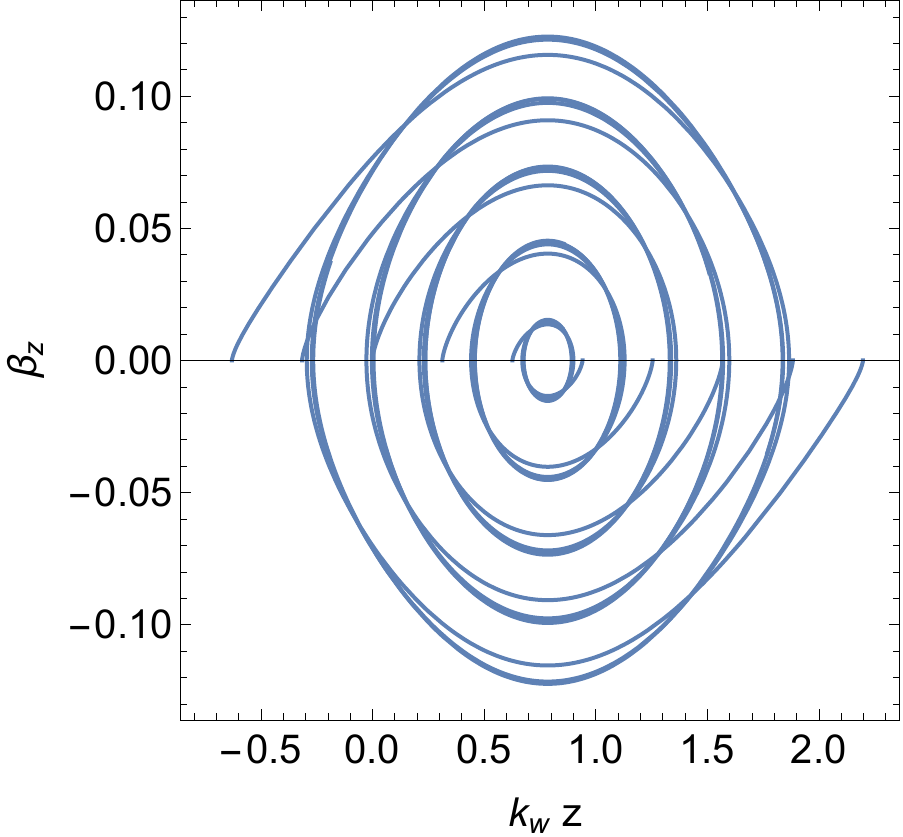}
\caption{ Left panel: Particle trajectories from numerical integration for particles in counter-propagating  linearly polarized wiggler and EM fields
(integration of Eq. (\protect\ref{1}) with $\eta_g = \eta_{EM,b} =0$.).
In this particular example $k_{w,b}=1$,  the switch-on function is $G=1 - e^{-t/\tau}$ with $\tau = 6\pi$, $\delta = 0.01 $. The motion consists of fast jitter and slow motion around the O-point in the $\beta_z - z$ plane. Initial locations of particles are clear seen at the $\beta_z=0$ line. Notice how particles are effectively ``pulled in'' towards the O-point  (by the ponderomotive force) -  this results in density enhancements. Compare with  Fig. \protect\ref{Opoint1}; only one island is pictured here; location of the O-point is not shifted as in  Fig. \protect\ref{Opoint}. Right panel:  same, but only with the slow term in  Eq. (\protect\ref{1}), $\propto \cos(2 k_{w,b}z) $}
\label{trajector1} 
\end{figure}

Equation of motion (\ref{01}) (with switch-on function re-instated) can be directly integrated. Assuming vanishing initial velocities and employing a switch-on function  $G=1 - e^{-t/\tau}$, and linearly polarized waves $\eta_g =\eta_{EM,b}=0$, 
particle trajectories, phase diagram and distribution are shown in Fig. \ref{trajector}-\ref{trajector1}.

Numerical integration  clearly shows parametric resonance (Fig.  \ref{trajector}, central panel) characteristic of the pendulum equation.  O-points  (density enhancements) are located at $ k_{w,b}z = \pi/4 + n \pi$, 
while X-points are at $ k_{w,b}z = 3\pi/4 + n \pi$   ($n=0,1...$). 
In case of adiabatic switching all trajectories are trapped (except those starting exactly at the X-points, a set of measure zero). 
Most importantly, direct numerical integration shows formation of density enhancements near the minimal of the ponderomotive potential.

Charge bunches  in the  beam  are created: Figures \ref{trajector}-\ref{trajector1} clearly demonstrate that an initially homogeneous distribution of charges is bunched by the ponderomotive potential.

To provide analytical estimates for the density distribution within a bunch, we note  that for a given parameter $\eta_\beta$ a probability to find particle at position $z$ is
\be
\frac{dp}{dz}  \propto \frac{1}{\beta_z}
\label{dpdz}
\ee
Assuming that initial  homogeneous distribution of particles over the phase translates in  homogeneous distribution in $\eta_\beta$ (this is approximately true, as can be verified by direct numerical integration), integration of Eq. (\ref{dpdz}) over $0< \eta_\beta<$  gives
\be
\int_0^1\frac{dp}{dz}  d  \eta_\beta = \ln \cot ( k_{w,b} z /2) \approx \ln \left( \frac{2}{k_{w,b} z} \right) 
\label{dpdz1}
\ee
where the last relation assumes $z \to 0$ (this is a shifted coordinate $z$). 

\begin{figure}[h!]
\centering
 \includegraphics[width=.99\textwidth]{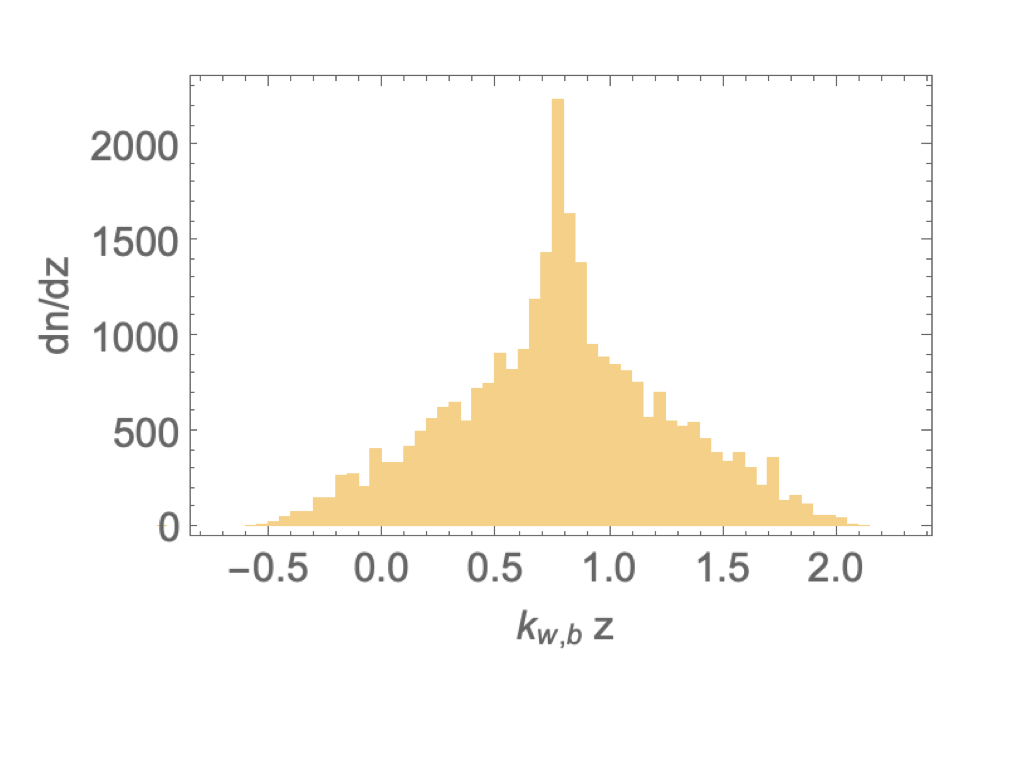}
\caption{Density distribution within a bunch, averaged over many periods of slow oscillations (arbitrary normalization, random initial phases).  The charge distribution  is logarithmically divergent, Eq. (\protect\ref{dpdz1}). }
\label{nave} 
\end{figure}

In Fig. \ref{nave} we zoom-in into the density distribution within a bunch, showing a nearly 2D charge distribution, consistent with mildly divergent distribution  (\ref{dpdz1})




\subsection{Non-harmonic wiggler}
The  proposed mechanism works, with some modifications, with non-harmonic wigglers as well. Naturally, the wiggler has to have large power  concentrated at the frequencies of the EM  to be amplified. 

\begin{figure}[h!]
\centering
 \includegraphics[width=.99\textwidth]{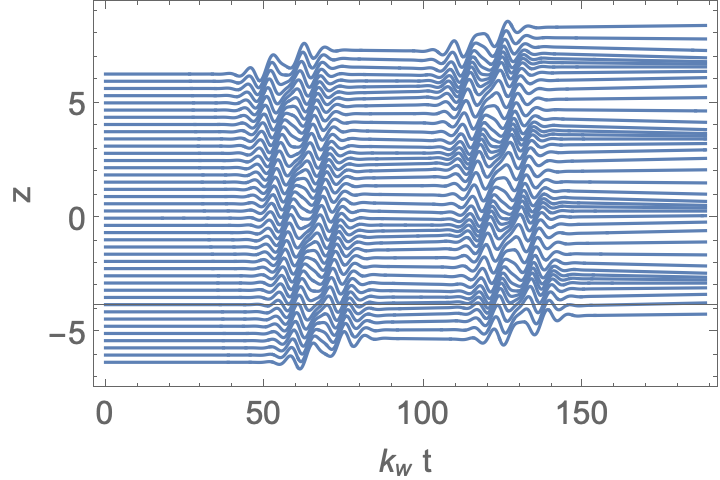}
\caption{ Particle trajectories  for  two gaussian wigglers located initially at  $z_{i,1} = 20 \pi$ and   $z_{i,2} = 40 \pi$; both wigglers have duration $\tau_G = 2 \pi$. First wiggler creates double density enhancements. The second wiggler both ''shakes''  them  producing coherent emission,  and further amplifies them.}
\label{double-gauss} 
\end{figure}

For example, consider  Gaussian  wiggler pulse(s) propagating to-the-left.
\be
B_{w,b} \propto a_{H,b} e^{- (t+z-z_i)/(2 \tau_G^2)}
\ee
where $z_i$ is the initial position of the wiggler's peak and $\tau_G$ is the duration. 
Strong effects are expected for short pulses,  $\tau_G \sim 2 \pi/ k_{w,b}$, see Fig. \ref{double-gauss}. Numerical integration clearly shows formation of charge bunches in the wiggler frame.

Thus, in a more complicated set-up, non-harmonic wigglers also produce bunches and coherent emission. We expects the effects of charge separation between the wigglers (when there is no ponderomotive force to keep  the charges separated), will be modified by plasma oscillations.

Conceptually, a non-harmonic wiggler pulse still creates charge enhancements. After the pulse propagated away, these charge enhancements oscillate electrostatically, so that the following  wiggler pulse  propagates through Langmuir turbulence, shaking the charged density enhancements and producing coherent emission.

 \subsection{Comparison with unmagnetized case}
Let us next  compare cases of {\it static} wigglers with no guide field and guide-field dominated case. Let in the frame of the wiggler the fluctuating field be $  B_w$ and wavelength $k_{w}$.  In this  frame a particle is moving with \Lf\ $\gamma_0$. (The case of \EM\ and static wigglers are nearly equivalent for $\gamma _0 \gg 1$, hence  we use the same notations for the static wiggler.)

A particle executes a trajectory  with longitudinal radius of curvature 
    \be
     R_c \sim \frac{ 1}{a_{H}   k_{w}}
     \label{Rc}
     \ee
It is much larger than the size of the transverse circle a particle makes, which for circular  polarized wiggler is  $r_g \sim a_{H}/k_{w}$.

Single particle emissivities (neglecting factors of the order of unity and effects of wiggler polarization.)
\ba && 
P_{\rm no-B} \approx  \frac{e^2}{c} \gamma_b^2 \om_{ B_w}^2
\nn &&
P_{\rm  B} \approx   \frac{e^2}{c}  a_{H}^{ 2}  \gamma_b^4  ( k_{w} c)^2  
\nn &&
\frac{P_{\rm  B}}{P_{\rm no-B}} = \left(  \gamma_b  \frac{ k_{w} c} {\om_B}\right)^2  \ll 1 
\nn &&
\om_{B_w} =\frac{e B_w}{m_e c}
\ea

  The total power  $P_{\rm  B}$ obeys the Larmor formula for curvature emission,  but the spectrum is different:  Landau-Pomeranchuk effects are dominant. 
  For a particle with \Lf\ $\gamma$ the radiation formation length is \citep{LLII}
\be
l_c \sim \gamma^2  \frac{c}{\om} \sim 1/ k_{w}
\ee
(relativistic electron separates by a wavelength from  own radiation for a time $\gamma^2$ times wave period.)  Thus, within  a radiation formation length the particle changes its velocity considerably. As a result, the emission is of Compton type, $\om \sim \gamma_b^2 k_{w} c$, not curvature (where we would have expected 
$\om \sim \gamma_b^3 {a_{H}   k_{w}} c$).

  In case of single particle motion, 
many  relations for the guide-field dominated case  can be recovered with a change 
 \be
 K_u \equiv  \frac{ e B_w}{m_e c^2 k_{w} } 
 \rightarrow a_{H,b} \gamma_b
 \label{KB}
 \ee
where $K_u$ is the  conventional undulator parameter. 
 

\subsection{Plasma  effects   in the beam and the background}
\label{plasma}

  The wiggler produces density fluctuation.
These  fluctuation of charge density will produce electrostatic field that will affect beam dynamics. We consider them next.  

 The particles move due to both electrostatic \Ef\ $E_z= - \partial_z \Phi$ ($\Phi$ is the electrostatic potential) and  the Lorentz force. Using relations (\ref{01}) for linearly polarized waves $\eta_g = \eta_{EM,b}= 0$, and adding electrostatic acceleration (in the non-relativistic regime $\beta_z \ll 1$), the axial equation of motion becomes
\be
\partial _t \beta_z = \partial_z \Phi + \delta  \left(  \cos \left(2 z
   k_{w,b}\right)- \cos \left(2 t k_{w,b}\right)\right)
   \label{40}
\ee

Using charge conservation
\be
\partial_t  \delta n + \partial_z \left( \beta_z( n_0+ \delta n )\right)=0,
 \label{41}
\ee
and Poisson equation 
\be
\Delta  \Phi = 4 \pi  e \delta n,
 \label{42}
\ee
the system (\ref{40}-\ref{42}) represent a closed set for charge density, electrostatic potential and velocity field.

Treating  density fluctuations $\delta n$ and $\beta_z'$ as small quantities, we find
\ba&&
\beta_z = - \delta  \frac{2 k_{w,b}} {(2 k_{w,b})^2 - \om_{p,b}^2} \sin (2 k_{w,b}t)
\nn &&
 \om_{p,b}^{2} =\frac{ 4 \pi  e^2 n_b}{m_e}
  \label{dnplasma}
 \ea
 where $ n_b$ is beam density in its frame. 
 
  Relation (\ref{dnplasma}) 
clearly demonstrates a resonance between Langmuir oscillations and  wiggler-induced jitter.
The resonance occurs when the double frequency of the wiggler (recall that the wiggler produces longitudinal oscillations at its double frequency) matches the plasma frequency. 

Importantly, under the corresponding approximation  of small amplitude of jitter motion,  the phase of the density oscillations is not affected, only the amplitude.
The two regimes have a name in the field of FEL research: (i) Compton regime $\om_{p,b} \ll k_{w,b}c$; (ii) Raman regime $\om_{p,b} \gg   k_{w,b}c$. As we discuss in \S \ref{Densityestimates}, in astrophysical setting the scattering may occur in either regime.

In conclusion, while  the amplitude of oscillations due to  the ponderomotive driving depends on the beam plasma frequency, its phase is not. Thus plasma effects in the beam affect the strength of driving, but  not the resonance condition.  

We leave numerical consideration of electrostatic effects in the beam to a future paper. The approach to follow  electrostatic oscillation in quasi one-dimensional approach for charged beams  has been previously outlined by  \cite{2005ApJ...631..456L,2010MNRAS.408.2092T}.

Presence of background plasma may affect operation of FEL in several ways: (i) plasma dispersion is modified, hence waves  are not vacuum waves;
(ii)  wave escape: emission should be either produced on a mode that evolves into vacuum mode and escapes directly, or should be converted into escaping waves; (iii) background plasma may compensate the charges bunches in the beam.

Let us address these issues in turn. Obliquely propagating modes in pulsar \mss, at frequencies below the cyclotron frequency, can  be separation in the X-mode (with \Ef\ perpendicular to the $\B-\k$ plane) \citep{1986ApJ...302..120A,Kaz91,1998MNRAS.293..447L,2019JPlPh..85d9008K}.
For parallel propagation we are left only with X-mode with near-vacuum dispersion
\be
\om_X \approx k c \left( 1- \frac{\om_{p, bulk} ^2}{\om_B^2} \right)
\ee
where $\om_{p, bulk}$ is the bulk (background) plasma density. In magnetically dominated plasma $\om_{p, bulk} / \om_B \ll 1$ the X-mode is nearly a vacuum mode.

 Most importantly, 
the X-mode  extends continuously from $\om \leq \om_{p,b}$ to $\om \geq \om_{p,b}$  \citep[\eg, Figure 2 in ][]{2007MNRAS.381.1190L}.  It does not suffer plasma/Landau absorption at $\om \sim \om_{p,b}$.
X-mode is escaping  mode: as plasma density decreases, it connects to the vacuum mode.

We leave consideration of 
possible 
effects of the  background plasma on the charge separation in the beam to a subsequent paper. Here we just note that 
though  background's plasma density is higher than that of the beam, in the beam frame the  background's plasma dynamics will be suppressed by $\sim \gamma_b^{-3/2}$.

\section{Generation of coherent emission}

\subsection{Growth rate of parametric instability}
Qualitatively, the period of slow oscillations (in the beam frame)
\ba && 
\tau_s \sim \frac{1}{k_{w,b}\beta_S} = \frac{1}{\sqrt{ k_{w,b}\delta}} = \frac{1}{\sqrt{a_{EM,b} a_{H,b}} } \frac{1}{\sqrt{k_{w,b} \om_B } }
\nn && 
\tau_s  \times (k_{w,b} c) = \left( \frac{1}{a_{H,b} a_{EM,b}} \right)^{1/2}  \left( \frac{k_{w,b} c}{ \om_B} \right)^{1/2} 
\ea
This is also a time to develop  charge bunches. 

The evolution (growth) of the amplitude of the \EM\ wave occurs on similar time scale. Qualitatively its evolution  is then  governed by
\be 
\partial_ t a_{EM,b} = \frac{a_{EM,b}}{\tau_s} = \sqrt{ a_{H,b} k_{w,b}\om_B} a_{EM,b}^{3/2}
\ee
with a solution
\ba && 
\frac{a_{EM,b}} {a_{EM,0}} = \frac{1}{(1- t \Gamma_b )^2}
\nn && 
\Gamma_b  = {\sqrt{a_{H,b}a_{EM,0} }} {\sqrt{k_{w,b} c \om_B}} /2
\label{Gamma}
\ea
where $a_{EM,0}$ is the initial amplitude of the EM wave.

Eq. (\ref{Gamma})  gives  the growth rate for parametric instability (micro-bunching)  due to the interaction of wiggler field with normalized amplitude $a_{H,b}$ and the \EM\ field with initial amplitude  $a_{EM,0}$ (both measured in the beam frame). 
Relation (\ref{Gamma}) shows two important points: (i) instability is explosive, reaching infinite amplitude in finite time; 
(ii) growth rate is determined by the initial value of the EM wave intensity $a_{EM,0}$.
Note that the growth rate is only mildly suppressed by the guiding \Bf, $\Gamma_b  \propto B_0^{-1/2}$.

\subsection{Growth rate in the  high-gain/SASE regime}

Growth rate of the parametric instability (\ref{Gamma}) depends on the initial amplitude of the \EM wave $a_{EM,0}$.
What determines $a_{EM,0}$? It is most likely determined by an external seed radiation. As a lower limit, we can use an  analogues of SASE (Self-Amplified Spontaneous Emission)  regime of FEL,  where 
micro-bunching is initiated by the spontaneous radiation.

Let us estimate how much spontaneous emission is generated by  plasma particles within  a  ``lethargy'' region $\sim r$ (measured in lab frame). In the beam frame the corresponding size is $\gamma_b r$.
In the beam frame the Poynting flux of a wiggler is 
\be
F_w = a_{H,b}^{2} \frac{B_0^2}{4\pi} c
\ee
The scattered flux from a layer of thickness of $ \gamma_b r $  is then
\ba &&
F_{scat} = n_b\sigma_M  \gamma_b r F _w 
\nn &&
\sigma_M = \left( \frac{k_{w,b} c }{\om_B} \right) ^2 \sigma_T= 4 \gamma_b^2 \left( \frac{k_{w} c }{\om_B} \right) ^2 \sigma_T
\label{sigmaM}
\ea
where $\sigma_M $ is magnetic cross-section for X-mode and $\sigma _T$ is Thomson cross-section, $n_b$ is the beam density in the beam's frame.
Equating  scattered flux $F_{scat}$ to the initial EM flux,
\be
F_{scat} = a_{EM,0} ^{ 2} \frac{B_0^2}{4\pi} c
\ee
we find
\be
 a_{EM,0}=  \sqrt{ r \gamma_b  n_b\sigma_T}  a_{H,b} \alpha
  \ee
  (recall: $\alpha = k_{w,b}c /\om_B$).
We find for temporal growth rate
  \be
  \Gamma _b \approx  \left( r n_b\sigma_T {\gamma_b}  \right)^{1/4} {  {a_{H,b}} }( k_{w,b}c )
\ee
In the observer frame
\ba && 
\Gamma = \frac{\Gamma_b}{\gamma_b}= \left( r n_b   \sigma_T  \right)^{1/4} \frac{  {a_{H}} }{{\gamma_b} } \om
\nn &&
\om= 2 \gamma_b c k_{w,b}= 4 \gamma_b^2 (c k_{w})
\nn &&
a_{H,b} = \gamma_b a_{H} 
\nn &&
n= \gamma_b n_b
\label{Gammaob}
\ea

Spacial growth length is $L_w \sim c/\Gamma$.
We can then define  the magnetic Pierce parameter $\rho_B$ as 
\ba &&
\rho_B= \frac{1/k_w}{L_w}  = a_H  \left( \gamma_b^ 2 \frac{J }{I_A } \frac{ r_e}{ k_w}   \right)^{1/4} 
\nn &&
J= n e c 
\nn &&
r_e =\frac{e^2}{m_e c^2}
\nn &&
I_A = \frac{m_e c^3}{e}
\ea
(dimensionless factors have been omitted in the definition of $\rho_B$). 
Different scaling of $\rho_B$  from the conventional Pierce parameter can be traced to  magnetic cross-section  (\ref{sigmaM}).

\subsection{Saturation of the parametric instability}

Calculations of the non-linear  saturation levels of coherent instabilities is a daunting task. It requires calculations of how non-linear back reaction from the emitted wave affects that particle distribution, and the condition when that back-reaction saturations. Any order-of-magnitude estimate should be taken with a grain of salt.

As a physically motivated assumption, we suggest the following criteria for the saturation level. The instability is a parametric excitation of EM waves. As the amplitude of EM waves growth (parameter $\delta$, Eq (\ref{01}), the energy of the motion of the trapped particles increases. The saturation will be reached, we assume, when the energy density of the motion of the trapped particles (of the initially cold beam)   approaches the energy density of the EM waves (in the beam frame). 

The ponderomotive potential (\ref{Vp}), with parameter $\delta$  increases linearly with EM wave intensity $a_{EM,b}$, while energy density of EM waves increases quadratically $\propto a_{EM^2}$. The balance is achieved at 
\be
m_e n_ b \frac{ \beta_S^2}{2} = a_{EM,b}^2 \frac {B_0^2} {4 \pi}  \to 
\frac{ a_{EM,b}}{a_{H,b}} = \frac{ \om_{p,b}^2}{ k_{w,b} c \om_B}
\ee
where typical velocity of trapped particles $\beta_S$ is given by (\ref{betaS}). This is an estimate of he saturation level of the EM waves in the beam frame. 

In the observer frame this gives 
\be
\frac{ a_{EM}}{a_{H}}  = {2}{\gamma}^2 \frac{  \om_p^2}{\om \om_B}
\ee
Or, in terms of conversion efficiency of the beam's energy into radiation $\eta_{b-r}$ ,
\be
\eta_{b-r}\equiv \frac{ B_{EM} ^2 /(4\pi)} { n \gamma m_e c^2} = 4 \gamma^3  a_H^4  \frac{\om_p^2}{\om^2}
\label{etabr}
\ee
(see Eq. (\ref{etabrE}) for numerical estimates in astrophysical applications.)

The corresponding  saturated velocity jitter  
\be
\beta_{0,Sat} = \sqrt{2} a_{H,b} \frac{\om_{p,b}} {k_{w,b}} = \frac{a_H}{ \sqrt{2 \gamma}} \frac{\om_p}{k_w c}
\label{betaS} 
\ee
This is also an estimate of the applicability of the cold beam approximation. If the velocity spread in the beam is larger that (\ref{betaS}), then the parametric instability is suppressed. 
As we discuss in \S \ref{thermal}, the velocity spread in the beam frame is much smaller than in the observer frame, so the constant  (\ref{betaS}) is not very strict. Also, in astrophysical applications we expect   ${\om_{p,b}} \sim  {k_{w,b}} $, \S \ref{Densityestimates}. 

\subsection{Simple emissivity model of a charged bunch}

As a simple estimate  we can assume that particles within each half a period are bunched into a charged layer. This charged layer is shaken by the wiggler and emits coherently.  The surface charge density  for each layer is
\be
\sigma_e = e n_b \frac{\pi}{k_{w,b}} = e n  \frac{\pi}{k_{w} }   , 
\ee
 it oscillates with velocity $a_{H,b}c$ in its frame. Using jump conditions for oscillating fields on the oscillating charged layer, 
 the resulting Poynting flux  and energy density of radiation are
 \ba &&
 P _b=  2 \pi a_{H,b}^{2} \sigma_e^{ 2} = 2 \pi^3 e^2 c \frac{  n ^{ 2}}{k_{w,b}^{ 2}} 
  a_{H,b}^{ 2} 
  \nn &&
  \epsilon_b  = \frac{P}{c}
 \ea
 (this is the value on  each side of a charged layer - total energy  loss is two times larger). 
 
 In the lab frame
 \be
 P  = P_b \gamma_b^2 = 2 \pi^3 \gamma_b^4 \frac{ e^2 c^3 n^{ 2}}{\om^2} a_{H}^{2}= 
 \frac{\pi^3}{8} \frac{ e^2 c n^{2}}{k_{w,b}^{2}} a_{H}^{2}
 \ee
  Scaling $\propto n^2$ clearly indicates a coherent process.

\subsection{Constructive interference from different bunches}

\begin{figure}[h!]
\centering
\includegraphics[width=.99\textwidth]{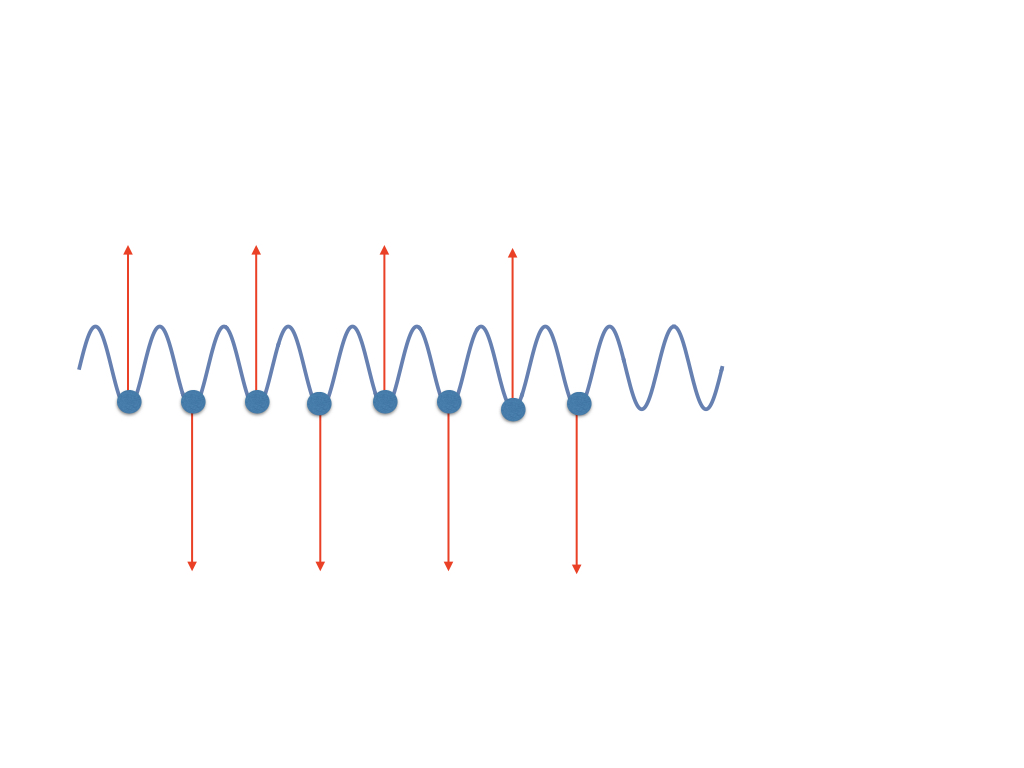} 
\caption{Constructive addition of emission from different charged bunches. Wiggled curve represent the ponderomotive potential. Particles are bunched at the minima of the potential creating charged bunches (solid circles). Bunches are shaken by the wiggler (red arrows). By the time an \EM\ wave from a give bunch arrives at the location of a neighboring one, the velocity reverses, so that waves add constructively. }
\label{bunches-consgructive} 
\end{figure}

The intensity  for parallel propagating emitted \EM\ wave is given by \citep[][Eq. 14.70]{jackson_99}
\ba &&
\frac{d^2 I}{d\om d\Omega} =
\frac{\om^2 }{4 \pi^2 c^3} \left| {\cal S} \right|^2
\nn && 
{\cal S} \propto  \sum_m \int dt \int dz j_x e^{i\om ( t - z_m)}
\ea
where the sum is over the location of the charges. Neglecting axial oscillations.
\be
\int dz \beta_x = 
(-1)^{m+1} \frac{1}{\sqrt{2}} \left( (a_{EM,b} -  a_{H,b}  )  \cos( k_{w,b}t) +(a_{EM,b} +  a_{H,b}  )  \sin( k_{w,b}t) \right) \approx
(-1)^{m}  a_{H,b}  \cos (\pi/4 +  k_{w,b}t) 
\ee 
where in the last relation we neglected $ a_{EM,b} \ll   a_{H,b} $. Thus, emission from different charged layers adds constructively, Fig. \ref{bunches-consgructive}

Thus, within the simple emission model, each layer emits coherently, and emission from different layers add constructively.

\section{Astrophysical viability}
\label{viability}

\subsection{Plasma parameters in pulsars and magnetars}
The suggested mechanism of coherent radio  production depends on two ingredients: the wiggler and the reconnection-generated beam. 
In \cite{2020arXiv200616029L} we discussed   properties of  firehose-excited wigglers, and 
  in \S  \ref{Thewiggler} we discuss   properties of  \Alfven  (low frequency)  wigglers. Next we consider the expected properties of the  particle beam and astrophysical applications, concentrating on   \Alfven  wigglers (scaling $k_{w,b}$ to the local radius, not the plasma properties).

Two scalings for the plasma density of the beam are viable: the pulsar-like \citep{GJ}  and magnetar-like  \citep{tlk}. As we expect the FEL to operate both in pulsars (Crab)  and magnetars/FRBs, we consider both cases in parallel.


Let's assume that radio emission is generated at distance $r$, and typical wiggler wavelength is related to $r$, but is somewhat  smaller,
\be
k_{w}= \eta_w \frac{2 \pi}{r}, \, \eta_ w \geq 1
\label{etaw}
\ee

For a beam of \Lf\ $\gamma_b $ the emitted frequency is
\be
\om  \sim 8\pi \eta_w   \gamma_b^2\frac{c}{r}
\label{om1}
\ee
The required \Lf\ is
\be
\gamma_b = \eta_w^{-1/2} \left( \frac{\nu r}{4 c}\right)^{1/2} =
\left\{
\begin{array}{c}
10^2 \,   \eta_w^{-1/2}   \left( \frac{r}{R_{NS}} \right)^{1/2} \nu_9^{1/2} \\
  10^3 \,  \eta_w^{-1/2}  \left( \frac{r}{R_{LC}} \right)^{1/2} \nu_9^{1/2}
\end{array}
\right.
\ee
where for two estimates the radius is  normalized to the \NS\ radius (top row) and Crab's \LC\ radius (bottom row). 

Both estimates  are very reasonable: such \Lf\ are indeed expected both in magnetar \mss\ \citep{2007ApJ...657..967B,2017SSRv..207..111C}  and in the reconnection current sheets in Crab pulsar \citep{2007ApJ...670..702Z,cerutti_15,2016JPlPh..82c6303C}.  {\it Thus, even the longest wiggler, with the wavelength of the order of the distance to the star, can reasonably produce GHz radio emission with relatively mild, by  pulsar standards, {\Lf}s.}

\subsection{Density estimates: Crab and magnetars}  
\label{Densityestimates}

For magnetars, 
as an estimate of the beam density (in lab frame) we can use
\be
n   = (\Delta \phi) \frac{B}{4 \pi e r}
\ee
where $ (\Delta \phi) $ is twist angle of magnetospheric field lines \citep{tlk}.
Using Lorentz transformation to the beam frame,
\ba &&
 k_{w,b}   = 2 \gamma_b k_{w} 
 \nn &&
 n_b = \frac{n }{ \gamma_b}
\ea
we  find the ratio of beam plasma frequency to the wiggler frequency (in the beam frame)
\be
\frac{\om_{p,b}}{ c k_{w,b}}  
\approx 
\left\{
\begin{array}{c}
 10^4  \times b_q^{1/2}  (\Delta \phi)^{1/2} \eta_w^{-1/4} \left( \frac{r}{R_{NS}} \right)^{-7/4} \nu_9^{-3/4}\\
 3  \times \left(  \frac{b_q}{0.1} \right) ^{1/2}  (\Delta \phi)^{1/2} \eta_w^{-1/4} \left( \frac{r}{R_{LC}} \right)^{-7/4} \nu_9^{-3/4}P_{0.03}^{-7/4}\\
  \end{array}
 \right.
\label{omm}
\ee
(For Crab $b_q \approx 0.1$.)
Though the factor in (\ref{omm}) is large, $\sim 10- 10^4$, all the parameters are smaller than unity,
$b_q^{1/2}, \,   (\Delta \phi)^{1/2},  \, \eta_w^{-1/4}, \,  \left( {r}/{R_{NS}} \right)^{-7/4} \leq 1$.
Thus, wiggler-plasma resonance effects (\S \ref{plasma}) can be important (so that density bunching is  enhanced). 

Alternatively, scaling beam density to the \cite{GJ} density
\ba &&
n_{GJ} = \frac{\Omega B}{2 \pi e c}
\nn &&
\frac{\om_{p,b}}{ c k_{w,b}'}  
\approx 
\left\{
\begin{array}{c}
 10^4  \times b_q^{1/2}  \eta_w^{-1/4} \left( \frac{r}{R_{NS}} \right)^{-5/4} \nu_9^{-3/4}P_{0.03} ^{-1/2}\\
 3 
  \times  \left(  \frac{b_q}{0.1} \right) ^{1/2}   \eta_w^{-1/4}\left( \frac{r}{R_{LC}} \right)^{-5/4} \nu_9^{-3/4}P_{0.03} ^{-7/4}
 \end{array}
 \right.
\label{omm1}
\ea
where \NS\ period is scaled to Crab.

This is a very important result: we find that   in magnetars perturbation of the \ms\  with scales somewhat smaller that the size of a \NS, as well as  in Crab pulsar perturbation of the \ms\  with scales somewhat smaller that the \LC,  are likely to produce oscillation of the beam plasma in resonance with the charge oscillations in the beam. In magnetars the required \Lf\ is $ \gamma \sim  10^2$, in Crab $ \gamma \sim 10^3$:  all reasonable estimates. 

  Possibility of wiggler-beam plasma resonance adds further complication. Resonant interaction enhances bunching, but it 
  depends sensitively on the properties of the driver and the dissipation processes; less so on the power of the driver.
  

\subsection{Growth rate: Crab pulsar and magnetars}
Relation (\ref{Gammaob}) give the growth rate of the parametric (bunching) instability. For astrophysical applications we chose two cases: Crab pulsar and magnetars, \S \ref{Densityestimates}.

Parametrizing wiggler wavelength by (\ref{etaw}), 
the condition 
\be
\frac{\Gamma   }{c/r}  \geq 1
\ee
requires
\be
\eta_w^{1/2} a_{H}  \geq \left( \frac{ c^2} {n \sigma_T \om^2 r^3}  \right)^{1/4} 
\label{growth}
\ee
This is a condition on the amplitude of the wiggler $a_{H} $ and its typical length $r/\eta_w$, so that in the SASE regime the  spacial growth rate is larger than the distance to the star.

Using density parameterizations of \S \ref{Densityestimates}, 
the condition (\ref{growth})  then gives
\ba && 
\eta_w^{1/2} a_{H}  \geq 6 \times 10^{-3} b_q ^{-1/4}\nu_9^{-1/2} (\Delta \phi)^{-1/4}  (r/R_{NS}) ^{-1/2}  \,  \mbox{for magnetars}
\nn && 
\eta_w^{1/2} a_{H}  \geq   3 \times 10^{-2}  \nu_9^{-1/2} \,  \mbox{for Crab},
\label{aHCrab}
\ea
 Thus, in both cases mild wiggler intensity $ a_{H}  \geq 10^{-3}$ is needed (recall that $\eta_w \geq 1$, Eq. (\ref{etaw})). 
 
 
\subsection{Expected brightness temperature}
     
     Estimating/calculating the power of coherent sources - in astrophysical setting when no lab technician is on-site - is, in some sense, a treacherous road.  Coherence depends on the subtle addition of phases of emitted wave; the saturation levels depend on  non-linear back-reaction  of coherently added waves on the kinetic properties of the  distribution function.

There are two ingredients for the production of radiation:  the wiggler     and the  beam. Comparing the expected energy densities in the beam and the wiggler
\be
\frac{ \gamma n   m_e c^2} {a_{H}^{ 2} B_0^2 /(8\pi)}
=
\left\{
 \begin{array}{c}
10^{-14}  \times b_q^{-1}  a_{H}^{-2} \eta_w^{-1/2} \left( \frac{r}{R_{NS}} \right)^{5/2} \nu_9^{1/2}  \\
6 \times 10^{-9}  \times  b_q^{-1}  a_{H}^{ -2}\eta_w^{-1/2} \left( \frac{r}{R_{LC}} \right)^{7/2}  \nu_9^{1/2} P_{0.03} ^{5/2}
\end{array}
\right.
\label{ah2}
\ee
for magnetar and pulsar scalings. This demonstrates that beam energy density is typically much lower than that of the wiggler: the beam cannot smooth out the wiggler field. 
 We conclude that for typically the energy density of wiggler's turbulence  is much higher than that of the beam. It is then the energy of the beam that determines the resulting radiation: energy in the wiggler is not a limiting factor.
     

The beam-radiation conversion efficiency (\ref{etabr}) evaluates to
\be
\eta_{b-r} = a_H^4 \gamma_3^3 \nu_9^{-2} \times
\left\{
\begin{array}{cc}
5 \times 10^4  & \mbox{Crab}
\\
 10^{15} \, b_q (\Delta \phi) \left( \frac{r}{R_{NS} }\right) & \mbox{magnetars}
\end{array}
\right.
\label{etabrE}
\ee
Values  of $\eta_{b-r} $ are both highly dependent on the amplitude of the wiggler $\propto a_H^4$, and have large numerical factors. This implies, first, that ver weak wigglers, with $a_H \ll 1$ are sufficient to convert a faction of the beam energy into radiation, and, second, that this conversion  efficiency is almost a threshold effect. 

 The resulting brightness temperature is  \citep{2019arXiv190103260L}
 \be
 {k_B {T}_b} = 
 \eta_{b-r}  (2\gamma)^{4}  \frac{  \lambda ^3  m_e c^2 n  }{2\pi} 
 \label{05}
 \ee
 where  $\lambda = c/\nu$ is wavelength. 
      
      Using two parameterizations for plasma density,  we find
      \be
      T_b =\eta_{b-r} \times \gamma_3^4\nu_9^{-3} 
      \left\{
      \begin{array}{cc}
       10^{43} {\rm K} \, \,  \,   b_q (\Delta \phi) \left( \frac{r}{R_{NS}} \right)^{-1} & \mbox{magnetars}
      \\
      2       \times 10^{33} {\rm K} & \mbox{Crab}

      \end{array}
      \right.
      \label{Tb} 
      \ee
      for magnetar and G-J scaling correspondingly. 
      
      Somewhat surprisingly (given the order-of-magnitude estimates) the brightness temperature estimates  (\ref{Tb}) match both the FRBs and Crab GPs 
\citep{1977puls.book.....M,Melrose00Review,2004ApJ...616..439S,2007Sci...318..777L,2019A&ARv..27....4P}. We consider this a major, and  unexpected, success of the model.


\subsection{Energetics}

Finally, let us comment on the energetics of FRBs. 
In the case of FRBs the energetics is constrained by the accompanying X-ray bursts  $E_X$ ({\it not} the FRB itself). 
The required size  $l_M$ of a region of dissipated magnetic energy is 
\be
l_M \sim 2 \pi^{1/3} \frac{E_X^{1/3} } {B_{NS} ^{2/3}} =
5 \times 10^4 {\rm cm} b_q^{-2/3}E_{X, 40}^{1/3}
\ee
about a football field for the quantum  critical surface \Bf.
Time-alignment of few msec between radio and X-rays  imply then that radio emission in FRBs  is also generated relatively close to   the NS's surface, at $r \leq 10^8$ cm.

\section{Advantages of guide-field dominate wiggler}

\begin{figure}[h!]
\centering
\includegraphics[width=.49\textwidth]{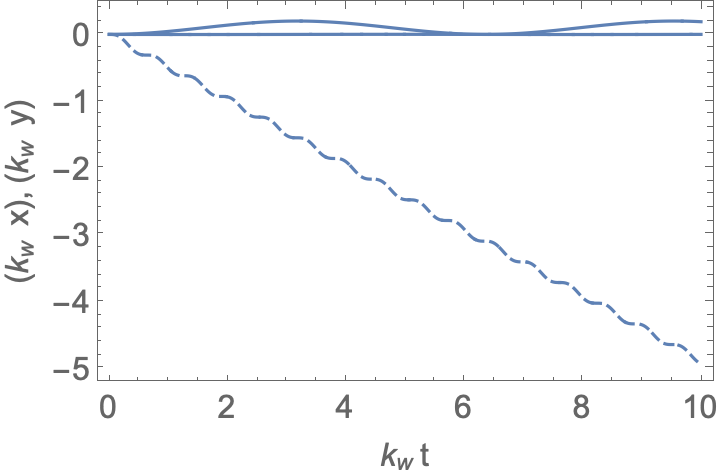} 
\includegraphics[width=.49\textwidth]{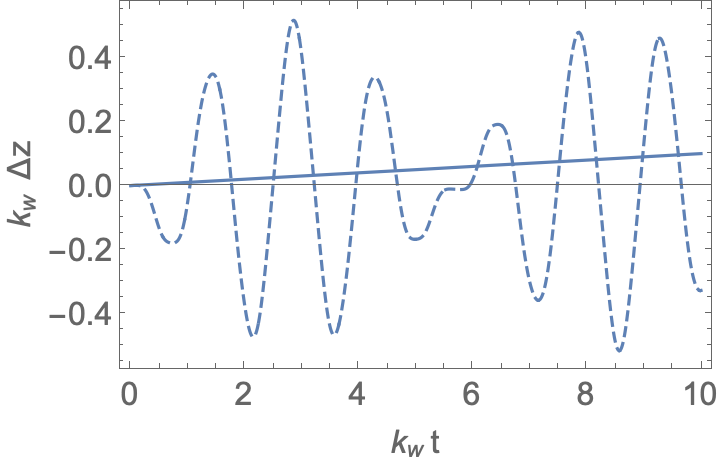} 
\caption{Comparison of cross-field drift and phase separation in wigglers without guide and in the  guide field dominated regime, static wigglers. 
Left pane:  cross-field displacement for guide-field dominated  wiggler $a_{H,b}=10$ (solid line) and no guide field (dashed lines for x and y) as a function of time.
Right panel: axial separation  $\Delta$ of two particles with  initial velocities $\beta_1 =0.98$ and $\beta_2=0.99$.  In the guide field dominated regime  phase coherence is much better preserved. }
\label{xy-drift} 
\end{figure}

In the present  astrophysical application guide-field dominates linear wiggler has a number of advantages. 
First, In laboratory FEL  linearly  polarized wigglers produce $\B_0 \times \nabla \delta\B$ drift that causes the beam to  drift away and expand. This is a fatal problem in the lab because the beam blows up.  The guide field suppresses drifts, Fig. \ref{xy-drift}. 

Most importantly, guide field dominance helps to maintain beam coherence, Fig. \ref{xy-drift}.b. Without the guide field particles with different energies follow different trajectories, and quickly lose coherence even for small initial velocity spread. In contrast, in the guide-field dominated regime all particles follow, basically, the same trajectory. Hence coherence is maintained as long as the velocity spread {\it in the beam frame} is $\Delta \beta \leq 1$.  

\section{Discussion}

We construct a model of coherent radio emission generation in  pulsars,  magnetars and  Fast Radio Bursts. We suggest that radio emission in all these cases is  reconnection-powered. In the case of Crab Giant Pulses,  the reconnection events  occur outside the light cylinder in the magnetic equator current  sheet, while in the case of  magnetars and FRBs it  occurs deep inside the \NS\ \mss. 

 
 The emission mechanism is a variant of  Free Electron Laser. In \NS's \mss\ the 
FEL operates in a guide-field dominated regime, $\om_B \gg \gamma_b k_{w} c$. This is highly  unusual regime by laboratory standards. 

Reconnection events launch fast particle beams propagating along \Bf\  perturbed  either by the pre-existing or self-generated  turbulence.  The turbulence may be driven  either by magnetohydrodynamical effects, or by the firehose/two stream instability of  counter-propagating plasma components. Turbulent fluctuations (the wiggler) create charge bunches in the beam (via ponderomotive force)  that coherently scatter the wiggler field.

The present model explains a number of fairly subtle features of radio emission in Crab, magnetars and FBRs:
\begin{itemize}
\item conforms with the predicted  nearly simultaneous observations of radio and X-rays burst in magnetars, with radio slightly ahead of the X-rays \citep{2020ApJ...898L..29M},  as predicted by \cite{2003MNRAS.346..540L,2017ApJ...838L..13L}.
\item operates in a very broad range of \NS's parameters: the model is {\it independent} of the value of the \Bf. It is thus applicable to a broad variety of NSs, from  fast spin/weak \Bf\ millisecond pulsars to slow spin/super-critical   \Bf\ in magnetars, and from regions near the surface up to (and a bit beyond of) the \LC.
\item the model require only mildly narrow distribution of beam's particles, $\Delta p /p_0 \leq 1$ and the spectrum of turbulence  $\Delta k_{w,b}/k_{w,b} \leq 1$
\item reproduces (multiple) emission bands seen in Crab and FRBs; the model also can produce   broader  spectrum emission.
\item matches the polarization properties of FRBs (those that show narrow emission bands are linearly polarized, while broad-band emission can be circularly polarized)
\item naturally gives correct estimates for the brightness temperatures both in pulsars and FRBs.
\item  The model is also consistent with overall duration of FRBs, from microseconds to milliseconds \citep{2020arXiv201005800N}: the dissipation size needed for a medium X-ray  flare  accompanying  an FRB is $\sim 100$  meters  $= 10^{-2}$ of the radius; hence FRB, emitting along B-field it might last $\sim  10^{-2}$ of the period - milliseconds, if the source works long enough. But  FRB can be as short as light crossing time over 100 meters, a microsecond. 
\end{itemize}

 We hypothesize that the radio emission is generated during the initial stage of magnetospheric reconnection, while the \ms\ is still relatively clean of the pair loading. The  giant $\gamma$-ray flare from the magnetar SGR 1806 - 20 had a rise time of only $200$ micro-seconds \citep{palmer}, matching the duration of the radio flare.
  In a possibly related study  of relativistic reconnection by \cite {2017JPlPh..83f6301L, 2017JPlPh..83f6302L,2018JPlPh..84b6301L}, it was found that in highly  magnetized plasma the reconnection process driven by large scale stresses (magnetically-driven collapse of an X-point) has an initial stage of extremely fast acceleration, yet low level of magnetic energy dissipation. This initial stage of reconnection may produce unstable particle distribution in the yet clean surrounding, not polluted by pair production.


Other points of importance include:
\begin{itemize}
\item  Radio emission is generated during the {\it initial} stages of magnetospheric reconnection, as argued by \citep[][]{2003MNRAS.346..540L,2016ApJ...824L..18L}.   \cite {2017JPlPh..83f6301L, 2017JPlPh..83f6302L,2018JPlPh..84b6301L} found that in highly  magnetized plasma the reconnection process driven by large scale stresses (magnetically-driven collapse of an X-point) has an initial stage of extremely fast acceleration, yet low level of magnetic energy dissipation. 
Observations of \cite{2020ApJ...898L..29M} confirm that radio leads high energy.
\item FRB duration and intrinsic time structure   is  determined by the lateral (not radial) extension/structure of the emission \citep{2020ApJ...889..135L}. Roughly speaking, if both the wiggler and the particle beam have extension $\sim R_{NS}$,  the effective radial direction of  the generated pulse would be $\sim c/( R_{NS} \gamma_b^2)$ - much shorter than the observed duration of an FRB. 
\citep[In passing we note that the closest analogue - Solar type-III bursts -  also sometimes show fine spectral structure, classified as type-IIIb ({\it stria}) bursts][we think  FRBs'  fine structure is different  in origin.]{1967AuJPh..20..583E}.
\item Large \Bfs\ at the FRB production cites are {\it required},  otherwise  coherently emitting particles will have dominant ``normal'' (synchrotron and inverse Compton) losses   \cite[]{2017ApJ...838L..13L,2019arXiv190103260L}. In high \Bf\ instead of large oscillations with momentum $p_\perp \sim a_A m_e c$, the coherently emitting particles experience mild  $E \times B$ drift. 
\item  The high energy and radio burst from SGR 1935+2154 lies not far from the so-called 
 G\"{u}del-Benz relationship, which relates the thermalized X-ray luminosity generated by magnetic reconnection in stellar flares to the nonthermal, incoherent, gyrosynchrotron radio emission. 
G\"{u}del-Benz correlation is usually interpreted that  first electrons are accelerated to non-thermal velocities and emit radio, then these particles are thermalized and emit thermal X-rays. The magnetar SGR 1935+2154 adds another point, with an important caveat that the radio emission in this case is coherent. (Interestingly, it also lies off to the ``expected'' side from the lower energetics fit: the radio is too bright, as expected for coherent emission.) 
Though the microphysics of this relation is far from clear, is it at least consistent with the concept that accelerating mechanism puts first energy into nonthermal particles that produce radio, and then that energy is thermalized, producing X-rays.
\end{itemize}

A number of principal issues need to be addressed:
\begin{itemize}
\item
reconnection and particle in magnetically dominated plasmas has been recently extensively studies \citep[\eg][]{2003ApJ...589..893L,2003MNRAS.346..540L,2005MNRAS.358..113L,2011SSRv..160...45U,2014ApJ...780....3U,guo_15a,2017JPlPh..83f6301L, 2017JPlPh..83f6302L,2018JPlPh..84b6301L,werner_17}. Acceleration in relativistic reconnection is complicated. It  depends on the details of plasma components and the value of the guide field  \citep[\eg competition of tearing mode and  drift kink modes][]{2003MNRAS.346..540L,2007MNRAS.374..415K,2007ApJ...670..702Z,2008ApJ...677..530Z,2014ApJ...783L..21S,2016MNRAS.462...48S}, as well as large scale properties of the magnetic  configurations  \citep{2017JPlPh..83f6301L, 2017JPlPh..83f6302L,2018JPlPh..84b6301L}.
 What is missing so far is similar studies in highly radiatively-dominated regime of magnetars.
\item The model relates magnetar/FRB emission to a particular type of pulsar radio emission, associated with Crab GPs.
Only few pulsars show a phenomenon of GP \citep{1968Sci...162.1481S,2001ApJ...557L..93R,2003ApJ...590L..95J,2004ApJ...616..439S,2007ApSS.308..563K}. Why only few pulsars show GPs?
\item  In case of Crab, effects of cyclotron resonance need to be investigated further (it is not realistic in magnetars, see \S \ref{Guidefielddominance}). Near the cyclotron resonance interaction of the  beam particles with the  wiggler/EM wave can be very efficient since the resonant scattering cross-section is huge \citep{1996ASSL..204.....Z,2006MNRAS.368..690L}.  
  This increases the efficiency of wiggler-beam coupling. Effects of radiative damping should be taken into account:
  \ba &&
  \tau_b \approx  \frac{m_e c^3}{e^2 \om_B^2}
  \nn &&
   \tau = \gamma_b \tau_b 
   \nn &&
   \frac{\tau \Omega}{2\pi} \approx 1 \times \gamma_3
   \ea
   where $  \tau_b$ is the cyclotron decay time in beam frame, $\tau$ is in the lab frame, and the numerical estimate is for Crab pulsar near the \LC. Thus, effects of  
 radiative damping, even at Crab's \LC,  may prevent operation of relativistic cyclotron masers \citep[where masing occurs due to orbital bunching of electrons][]{1974ApPhL..25..377S,1995PhRvE..52..998N}. 
\item  Plasma effects in the beam.   
  The wiggler produces density fluctuation. 
These  fluctuation of charge density will produce electrostatic field that will affect beam dynamics. In the present approach we considered what is called the Compton regime of FEL - neglecting beam plasma effects. This requires that wiggler/EM frequency (in the beam frame) is much larger than that the plasma frequency of the beam. In the opposite regime (Raman) the scattering in done not by a simple particles, but by plasma oscillations. It is expected that  the amplitude of oscillations due to  the ponderomotive driving depends on the beam plasma frequency, but  its phase does not. Thus plasma effects in the beam affect the strength of driving, but  not the resonance condition.  
We leave numerical consideration of electrostatic effects in the beam to a future paper. The approach to follow  electrostatic oscillation in quasi one-dimensional approach for charged beams  has been previously outlined by  \cite{2005ApJ...631..456L,2010MNRAS.408.2092T}.
\item What are the effects of the background plasma?
Presence of background plasma may affect operation of FEL in several ways: (i) plasma dispersion is modified, hence waves  are not vacuum waves;
(ii)  wave escape: emission should be either produced on a mode that evolves into vacuum mode and escapes directly, or should be converted into escaping waves; (iii) background plasma may compensate the charges bunches in the beam. We leave consideration of 
possible 
effects of the  background plasma on the charge separation in the beam to a subsequent paper. Here we just note that 
though  background's plasma density is higher than that of the beam, in the beam frame the  background's plasma dynamics will be suppressed by $\sim \gamma_b^{-3/2}$.
\item Angular spreading and resulting coherence degradation.   Since wiggles are non-relativistic the coherence conditions are not affected by transverse wiggling. Internal beam spreading,  effects of ``emittance'' using laboratory FEL terminology, are better studies in full PIC simulations.
\item In the case of FRBs, it is not clear why repeaters typically show narrow emission bands with linear polarization, while  (apparent) non-repeaters  are broadband  with more varies polarization properties  \citep{2019A&ARv..27....4P}.
\end{itemize}


In conclusion, we developed a conceptually  new model for the generation of coherent emission in pulsars (Crab in particular), magnetars and FRBs. The emission is not rotationally, but reconnection-driven. A combination of analytical and (fairly basic) numerical results explain a surprisingly wide range of phenomena  (\eg emission stripe(s), brightness temperatures and polarization correlations), the model is fairly robust to wiggler/beam parameters (requires only mildly narrow distributions) and is {\it independent} of the value of the \Bf\ (hence applicable to a broad range of astrophysical objects). We encourage more detailed  analysis, especially using PIC simulations.

We would like to thank Roger Blandford,  Samuel Gralla, Igor Kostyukov,  Henry Freund, Amir Levinson, Mikhail Medvedev,  Alexander Philippov, Sergey Ryzhkov, Anatoly Spitkovsky.  {\it Python}   code was written by Yegor Lyutikov. We also thank him for comments on the manuscript. This work was initiated while ML was a graduate student at Caltech; discussions with Peter Goldreich are acknowledged.

  This work had been supported by 
NASA grants 80NSSC17K0757 and 80NSSC20K0910,   NSF grants  1903332 and 1908590.

 \bibliographystyle{apj} 
  \bibliography{/Users/maxim/Home/Research/BibTex}

  \appendix

\section{Alfv\'{e}n force-free solitons}
\label{Thewiggler}

As we discussed above, for highly relativistic particle  the difference between a static wiggler and a propagating EM packet of \Alfven waves is minimal
\citep[for analysis of linear waves in pulsar \mss\ see][]{1986ApJ...302..120A,1998MNRAS.293..447L,2019JPlPh..85d9008K}. 
Two types of wigglers/ \Alfven waves can be produced in the \mss\ of \NSs. First, large scale magnetospheric motions, associated with (pre-)flare global evolution of \Bfs\ may/will generate \Alfven waves with the typical wavenumber related to the local radius $r$, $k_{w,b}\sim \eta_ w (1/r)$, $\eta_ w \geq 1$. 
 Second, development of current driven instabilities, of the firehose-type, can lead to generation of \Alfven waves \citep[][also, Lyutikov \& Philippov, in prep.]{2020arXiv200616029L}. In both cases we expect  Alfvenic perturbations propagating in the \ms. In the present Chapter we consider properties of the nonlinear modes in the pulsar \ms 
 \citep[force-free modes have been considered by ][]{1998PhRvD..57.3219T,Gruzinov99,2002MNRAS.336..759K,2013arXiv1307.7782P, 2011PhRvD..83l4035L,2014MNRAS.445.2500G,2015PhRvD..92d3002G}

\subsection{Light darts}
\cite{2015PhRvD..92d3002G} found a number of solution for nonlinear  force-free perturbations. Here we extend their solutions to the problem of wiggler fields in pulsar \mss.
Consider force-free plasma in \Bf\ of value $B_0$ directed along the $z$-axis, subject to an  \EM\ perturbation.
First, consider perturbation in Cartesian coordinates,
\ba &&
{\bf A} =\left( A_{x}  (x,y, \xi_+)  {\bf e}_x +  A_{y}  (x,y,\xi_+)  {\bf e}_y +  A_{z}(x,y,\xi_+)  {\bf e}_z \right)  
\nn &&
{\xi_+}= k_z z- \om t
\label{AA1}
\ea
Ideal condition then requires $A_z=0$. 

Two remaining modes can be separated. The transverse components of the vector potential can be separated into curl-free (``O-modes'')  and div-free  ``X-modes'' components.

For ``O-modes'' the  transverse vector potential is a gradient of a function 
\ba &&
{\bf A}_\perp = \nabla _2 \Phi(x,y) = \partial_x \Phi(x,y) {\bf e}_x +  \partial_y \Phi(x,y) {\bf e}_y
\nn &&
 \om=k_z
 \nn &&
 j_z= k_z^2 \Delta_ 2 \partial_{{\xi_+}} \Phi(x,y,\xi_+)
 \nn &&
  \nabla_2 = \{ \partial_x, \partial_y\}
 \nn &&
  \Delta_ 2 =  \partial_x^2 +  \partial_y^2
  \label{11}
 \ea
 The force-free equations are satisfied for arbitrary $\Phi(x,y)$. These are ``light darts'' \citep{2015PhRvD..92d3002G}. They are fully nonlinear solutions of force-free equations. (Importantly, $j_z$ is a {\it linear}  function of  $ \Phi$.
In particular, if $\Delta_ 2\Phi=0$, there is no current and the perturbation becomes vacuum-like.
In cartesian coordinates harmonic 2D functions are divergent  (\eg\ $\propto \cos( k_1 x) \cosh(k_1 y)$), 
but this property will allow us to find new non-linear solution in cylindrical geometry, Eq. (\ref{ATE}).

Thus,  O-modes (as well as TEM modes in cylindrical geometry)  could be called force-free \Alfven wave. They carry energy only along $z$-axis. For example, for harmonic $\Phi(x,y) \propto e^{i (k_x x + k_ y y)}$ the Poynting flux averaged over a period is $\propto (k_x^2+k_y^2) k_z^2  {\bf e}_z$.

Second, there is a set of ``X-modes'', where the vector potential is 
\be
{\bf A}_\perp =  \nabla_2 \Psi(x,y,\xi_+) \times {\bf e_z} =  \partial_y \Psi(x,y,\xi_+) {\bf e}_x -  \partial_x \Psi(x,y,\xi_+) {\bf e}_y
\ee
This is just a  vacuum X-mode, $\J = 0$.

Let us generalize previous relations to cylindrical coordinates,  with perturbations of the  type 
\be
{\bf A} =\left( A_{TEM}  (r ,k_z z- \om t)  {\bf e}_r +  A_{TE}  (r,k_z z- \om t)  {\bf e}_\phi +  A_{TM}(r,k_z z- \om t)  {\bf e}_z \right)   e^{i m \phi}
\label{AA2}
\ee
(using the standard notation for cylindrical wave-guide modes: Transverse Magnetic (TM), Transverse Electric (TE) and Transverse electromagnetic wave (TEM) modes. 

In vacuum, radial and $k_z z- \om t$ dependance  is separable, while $k_z z- \om t$ dependance   is naturally required to be harmonic.
Propagating modes with $\om \neq 0$ then require: (i) TEM mode requires $m=0$, $\om=k_z$, $A_{TEM} \propto 1/r$ (one needs  a cylindrical surfaces such as a coaxial cable to support a TEM wave); (ii)   TE mode requires $m=0$,  $A_{TE} = J_1\left( \sqrt{\om^2 - k_z^2} r \right) $; TM mode requires $k_z=0$, $ A_{TM}= J_m(r \om)$;  

In force-free, the ideal condition requires that the TM mode must have $\om=0$, $k_z =0$,  $A_{TM} \propto r^{\pm m}$: only static, $z$-independent and limited in radius solution.  Transverse Magnetic waves do not exist in force-free plasma.

 For the TEM mode, we find that for  $m=0$ and $\om=k_z$ any solution 
is   fully  {\it non-linear} with arbitrary $A_{TEM} \left[(r, k_z(\om-t) \right]$. 
\be
{\bf A}_{TEM} =  A_{TEM} (r, k_z(z-t)  {\bf e}_r, \, m=0
\ee
("Light darts" are not limited to axially-symmetric perturbations,  as discussed above in Cartesian coordinates.)

TEM mode carries non-zero axial current (\cf, ({\ref{11}))
\be 
J_{z, TEM} = \frac{ k_z \partial_r \partial_{\xi_+}\left( r  A_{TEM} (r, \xi_+) \right)}{r}\equiv  k_z\partial_{\xi_+} ( \nabla_2 \cdot {\bf A}_{TEM} )
\ee

Finally
\be
{\bf A}_{TE} =J_1(r\sqrt{ \om^2 -k_z^2}) e^{i (k_z z- \om t)} {\bf e}_\phi
\label{ATE}
\ee
The mode ${\bf A}_{TE}$ is the vacuum TE mode, that has no associated current or charge. It satisfied the condition $ \nabla_2 \cdot {\bf A}_{TE}$.  It is a axisymmetric  analogue of the harmonic X-mode. 

(Regarding the terminology, we can these solutions solitons, but they are not, in   a conventional sense  - balance between dispersion and nonlinearity -  these are soliton-looking  fully nonlinear solutions.)

To summarize, in the present  treatment, there is an interesting correspondence: in vacuum there are  two modes, TM and TE; the TEM mode is discarded since it is divergent on the axis. In force free, TE mode remains, since it has zero current, TM mode is discarded, and  - somewhat  surprisingly - the  TEM mode is not divergent.

\subsection{Properties of force-free solitons}
The mode ${\bf A}_{TEM}$ describes an axially-symmetric  EM perturbation, that can be  limited both in $z-t$ and $r$,   an \Alfven soliton. For example, choosing (similar procedure can be repeated in Cartesian coordinates giving ``\Alfven sheets'')
\ba &&
A_{TEM} =  B_w r e^{-(r/r_0)^2} e^{- \xi_+^2}
\nn &&
{\xi_-}= k_z ( z-t)
\ea
($d_w$ is a $z$-scale of soliton in terms of $2\pi/k_z$)
we find fields and currents, see Fig. 
\ba &&
\B= -{2 {\xi_+}   (k_z r) e^{-{{\xi_+}^2}-\frac{r^2}{r_0^2}}}  B_w  {\bf e}_\phi + B_0  {\bf e}_z
   \nn &&
  E_r = B_\phi
  \nn &&
  \J = - {4 {\xi_+}} \left(1-\frac{r^2}{r_0^2}\right) B_w k_z e^{-{{\xi_+}^2}-\frac{r^2}{r_0^2}} {\bf e}_z
   \nn &&
   \rho = J_z
   \nn &&
   \B\cdot \E =0
   \nn &&
   B^2 -E^2=B_0^2 \geq 0
   \label{solion}
   \ea
  At every point the four-current is null, $|\rho| = |j|$.
  
  \begin{figure}[h!]
\includegraphics[width=.49\textwidth]{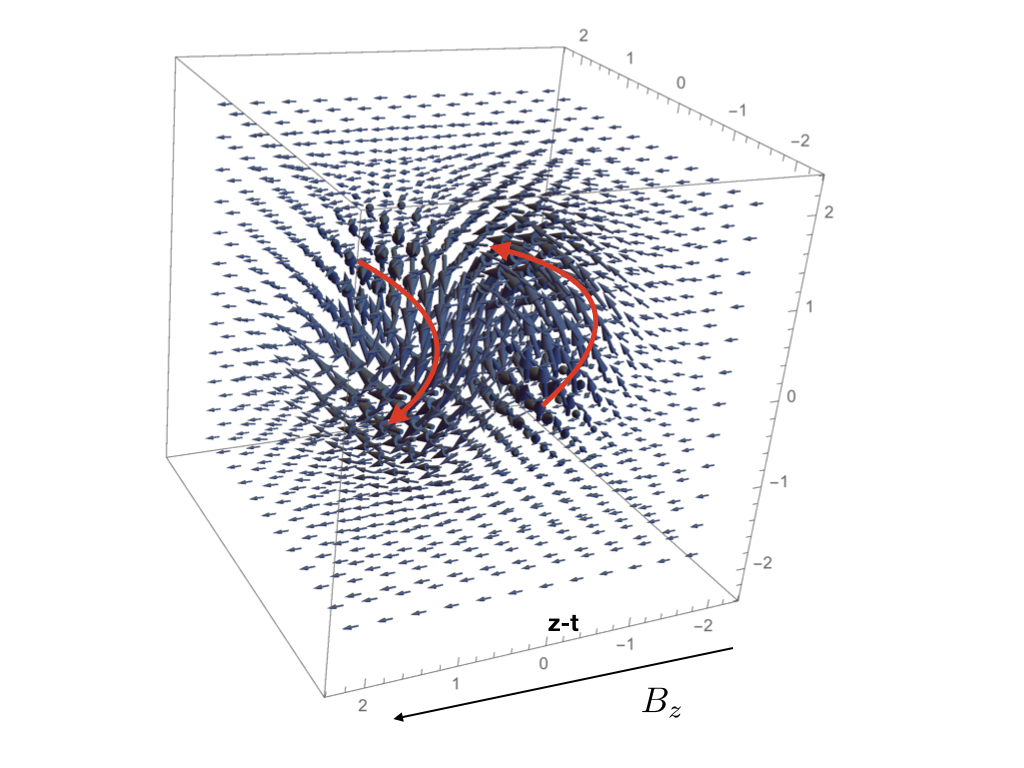}
\includegraphics[width=.49\textwidth]{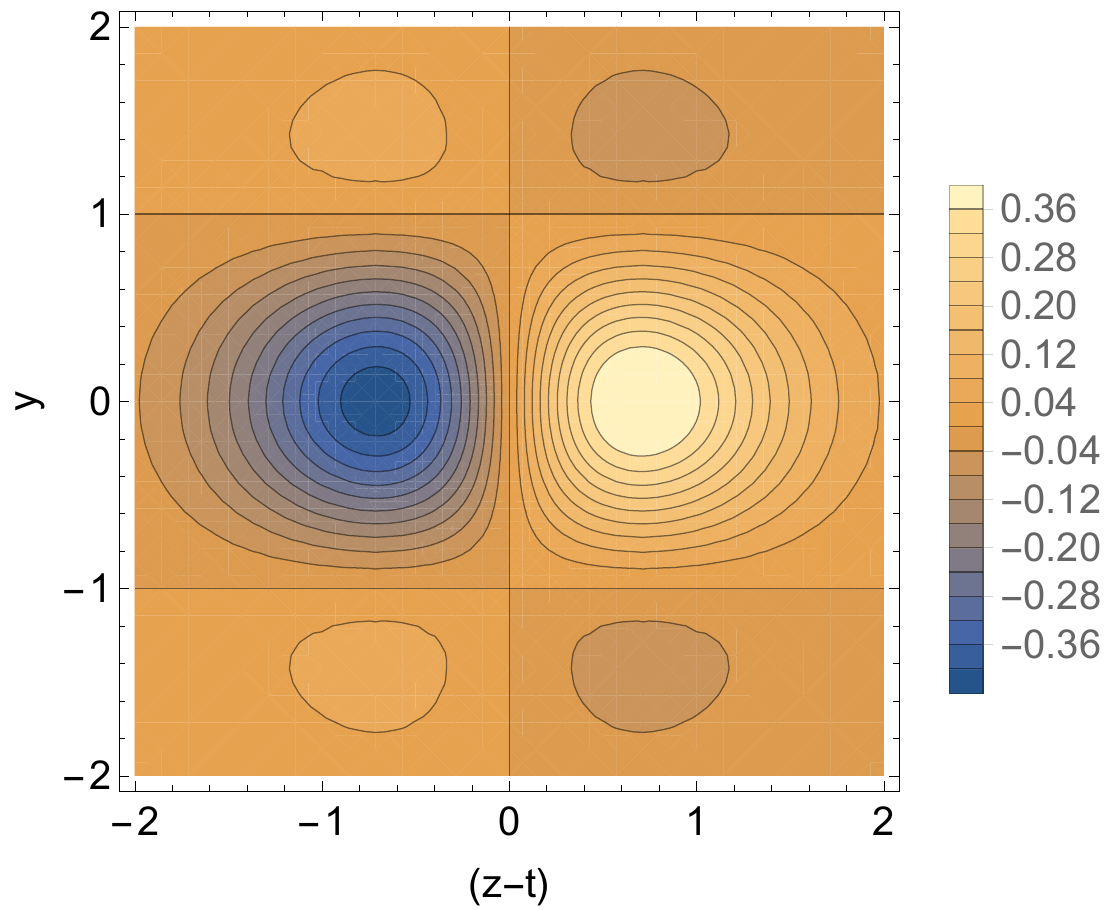}
\caption{ {\bf  Left  Panel: 3D rendering of \Bf\ in a non-linear  force-free \Alfven soliton, Eq. (\protect\ref{solion}). Amplitude of soliton-related \Bf\  perturbations is exaggerated for clarity.} Right Panel: Charge density (slice in the  $(z-t) - y$ plane). The  EM fields   of the soliton are limited in both transverse and longitudinal direction (double Gaussian). }
\label{rhoSoliton} 
\end{figure}

  Electromagnetic velocity (this is a drift velocity {\it across} \Bf)
  \ba &&
  \frac{ \E \times \B}{B^2} =
2 \left(  {\xi_+} k_z r a_{H,b}  e^{- \left( {\xi_+}^2+\frac{r^2}{r_0^2} \right) }  {\bf e}_\phi + 2 {\xi_+}^2 (k_z r) ^2 a_{H,b}^{2}    e^{-2 \left( {\xi_+}^2+\frac{r^2}{r_0^2}  \right)}  {\bf e}_z \right)
\left(1+4 {\xi_+}^2  ( r  k_z)^2 a_{H,b}^{2} e^{- 2 \left(  {\xi_+}^2+ \frac{  r^2}{r_0^2} \right) }\right)^{-1}
\approx
\nn &&
2  {\xi_+} k_z r a_{H,b}  e^{- \left( {\xi_+}^2+\frac{r^2}{r_0^2} \right) }  {\bf e}_\phi + 4  {\xi_+}^2 (k_z r) ^2 a_{H,b}^{2}    e^{-2 \left( {\xi_+}^2+\frac{r^2}{r_0^2}  \right)}  {\bf e}_z 
 \ea
 where $a_{H,b} = \delta B/B_0$ (in lab frame - this is different from the notation in the main paper.) 
  As the soliton propagates, it induces plasma rotation that changes sign in the middle, plus plasma motion along the $z$-direction.
  At $r \leq r_0$ the plasma rotates with nearly constant angular velocity.
    Another curious property of the solution is that the radial component of $\curl \B$ is exactly cancelled by $\partial_ t \E$.
  
   The total charge carried by the soliton is zero.   It is due to the cancellation of two pair of charges, at ${\xi_+}>,< 0$ and $ r>,< r_0$, Each of the total value $\pm (2 \pi /e^1) B_w r_0^2$. 
   Thus, the solution carries typical charge density 
   \be
   \rho_s \approx  \frac{B_w k_z }{ {2 \pi} }
      \label{rhos}
   \ee
   
   Comparing charge density (\ref{rhos}) to the GJ density $\rho_{GJ}$, and a density expected in \mss\ of magnetars
   \ba &&
   \frac{\rho_s }{\rho_{GJ} }\approx { a_{H,b} } ( {k_z R_{LC}}) \approx \frac{a_{H,b}}{\gamma_b^2} \frac{ R_{LC}}{\lambda}
   \nn &&
   \frac{\rho_s }{ {(\Delta \phi)} B_0 /r}=  \frac {a_{H,b}  }{{(\Delta \phi)}}  k_z r\approx \frac {a_{H,b}  }{{(\Delta \phi)  \gamma_b^2 }} \frac{r}{\lambda}
   \ea
where in the latter relations we used $\omega \sim c/\lambda \sim  \gamma_b^2  k_{w,b}c$.

Since the background plasma is expected to have total plasma density larger than the minimal one by a factor $\kappa \sim  10^3$, conditions of charge starvation  are
\ba &&
a_{H,b} \leq \gamma_b^2 \kappa \frac{\lambda}{R_{LC}} =  7  \gamma_{0,3} ^3 \kappa_3 
\nn &&
a_{H,b} \leq \gamma_b^2 \kappa(\Delta \phi)\frac{\lambda}{r} = 10^3  \gamma_{0,3} ^3 \kappa_3 (\Delta \phi)
\ea
for emission at $\lambda = 1 $ cm. Thus, charge starvation is  not likely to affect Alfven solitons.


\section{Momentum spread of the beam}
\label{thermal}

Here we demonstrate that (i) a mild spread  of particle momenta in the observer frame, $\Delta p/p \leq 1$ is sufficient to keep coherence; (ii) mild ``normal'' radiative losses  (non-coherent) help greatly in reducing the momentum spread of the particles.

\subsection{Relativistic kinematic reduction of  thermal  spread of the beam}

Let's assume  a fast beam  propagates with \Lf\ $\gamma_b$ and has thermal spread $\theta$ in its rest frame.
Juttner-Maxwell distribution for the beam in the observer frame is
\ba &&
f (p)=  \frac{1}{2 \theta \gamma_b  K_1(1/\theta)} e^{ - ( \gamma  \gamma_b  - p p_b)/\theta}
\nn &&
\gamma  =\sqrt{p^2+1}
\nn &&
p_b = \sqrt{ \gamma_b^2-1}
\nn &&
\int f dp = 1
\ea

Assuming  $1\ll \theta \leq \gamma_b$,
\be
f= 0.24 \frac{1}{ \theta \Gamma  K_1(1/\theta)} e^{ -(\gamma-\gamma_b)^2/(2 \gamma_b^2 \theta)} 
\label{fb}
\ee
The spread in the momentum (and the  \Lf)  of the beam in the observer frame is $ \Delta \gamma \approx  \gamma_b^2 \theta$. Inversely, a spread in the center of momentum frame of the beam is $\theta  = (\Delta  \gamma_b /  \gamma_b) / \gamma_b \ll  (\Delta  \gamma_b /  \gamma_b)  $.

Thus, for any relativistic beam with $\gamma_b \gg 1$, if  the  relative energy spread in the observer frame is  $ (\Delta  \gamma /  \gamma_b)  \leq 1$,  then the corresponding  spread in the frame of the beam becomes tiny $\ll 1$.


\subsection{Reduction  of  thermal  spread  due to cooling}

Particles  accelerated at reconnection will experience cooling via synchrotron, IC and curvature emission. As discussed  by \cite{1999MNRAS.305..338L} for curvature emission this  will lead to drastic reduction in the spread of the Lorentz factors in the beam frame. Generally,  if  cooling  scales as  
\be
\partial \gamma = -  c_1 \gamma^{1+\alpha}
\ee
  ($\alpha =1 $ for IC and  synchrotron cooling and $\alpha =3 $ for curvature),  the energy of  each particle evolves according to
\be
\frac{\gamma(t) }{\gamma_b} =  \left(1+ c_1 \alpha \gamma_b^\alpha t \right)^{-1/\alpha}
\ee
Integrating along trajectories we find that a spread in  Lorentz factors evolves according to
\ba &&
\frac{ \Delta \gamma}{\Delta \gamma_b} = \left(\frac{\gamma_b(t) }{\gamma_b} \right)^{2(1+\alpha)} 
\nn &&
\frac{ \Delta \gamma}{\gamma_b(t) } = \left(\frac{\gamma_b(t) }{\gamma_b} \right)^{1+ 2\alpha }  \frac{ \Delta \gamma_b}{\gamma_b}
\ea
where $\gamma_b(t) $ is the average \Lf\ at time $t$. Overall cooling of the beam by a factor of 2 reduces its  \Lf\ spread by a factor 
$2^   {1+ 2\alpha }  = 8, \, 128$ for $\alpha =1,3$. Thus, mild  overall cooling of the beam particles results in drastic reduction of the internal  spread of   Lorentz factors.

 \end{document}